\documentclass[5p,times]{elsarticle}

\journal{Future Generation Computer Systems}

\usepackage{graphicx}
\usepackage{algorithmic}
\usepackage{array}
\usepackage[caption=false,font=footnotesize,labelfont=sf,textfont=sf]{subfig}

\usepackage{url}
\usepackage[colorlinks]{hyperref}
\usepackage{soul}
\usepackage{amsmath,amssymb,amsfonts}
\usepackage{booktabs}
\usepackage{multirow}
\usepackage[table,xcdraw]{xcolor}
\usepackage{lipsum}
\usepackage{setspace}
\usepackage{tikz}
\usepackage{scalerel}

\newcolumntype{L}[1]{>{\raggedright\arraybackslash}m{#1}}
\newcolumntype{C}[1]{>{\centering\arraybackslash}m{#1}}
\newcolumntype{R}[1]{>{\raggedleft\arraybackslash}m{#1}}

\usepackage{ragged2e}

\definecolor{arylideyellow}{rgb}{0.91, 0.84, 0.42}
\definecolor{blond}{rgb}{0.98, 0.94, 0.75}

\newcommand{\tikztriangleright}[1][red,fill=red]{\scalerel*{\tikz \draw[rounded corners=0.1pt,#1] (0,-2.5pt)--++(0,5pt)--++(-30:5pt)--cycle;}{\triangleright}}

\newcommand{\lmttfont}{\fontfamily{lmtt}\selectfont}

\usepackage{pdfcomment}
\usepackage{todonotes}
\newcommand{\revision}[1]{{\textcolor{black} {#1}}}

\usepackage[most]{tcolorbox}

\tcbset{
    enhanced,
    breakable,
    left=2mm,
    right=2mm,
    attach boxed title to top left={
        xshift=0.5cm,
        yshift= -3.5mm
    },
    top=4mm,
    colback=blue!5!white,
    colframe=blue!75!black,
}

\newenvironment{mybox}[1]{%
    \begin{tcolorbox}[title={#1}]%
    \setstretch{0.95}}{
    \end{tcolorbox}
}

\begin{document}

\begin{frontmatter}

\title{Virtualizing Mixed-Criticality Systems: A Survey on Industrial Trends and Issues}

\author{Marcello Cinque}
\author{Domenico Cotroneo}
\author{Luigi De Simone*}
\author{Stefano Rosiello}

\address{DIETI - Università degli Studi di Napoli Federico II, Via Claudio 21, 80125 Napoli, Italy}



\cortext[correspondingauthor]{Corresponding author\\Email address: luigi.desimone@unina.it (Luigi De Simone)}

\begin{abstract}

Virtualization is gaining attraction in the industry as it promises a flexible way to integrate, manage, and re-use heterogeneous software components with mixed-criticality levels, on a shared hardware platform, while obtaining isolation guarantees. 
This work surveys the state-of-the-practice of real-time virtualization technologies by discussing common issues in the industry. In particular, we analyze how different virtualization approaches and solutions can impact isolation guarantees and testing/certification activities, and how they deal with dependability challenges. The aim is to highlight current industry trends and support industrial practitioners to choose the most suitable solution according to their application domains.

\end{abstract}


\end{frontmatter}



\section{Introduction}
\label{sec:introduction}





In recent years, we are witnessing the increasing adoption of virtualization technologies in industrial domains, such as railways, avionic, automotive, Industrial Internet of Things (IIoT), but also in telco systems with the recent development of 5G. \cite{shift2rail,windriveriot,hercules2020,5gcity_H2020,hermann2016design,klingensmith2019using, klingensmith2018hermes}. 
In such industrial domains, it is quite common to deal with so-called mixed-criticality systems, which integrate functionalities of different safety and/or time-critical into a common platform to reduce the size, weight, power, and cost of hardware. The integration of functionalities with different safety and time requirements leads to numerous challenges, especially when adopting virtualization technologies.

Even though virtualization easily supports mixed-criticality compositions since it implicitly provides software support for partitioning and running tasks on heterogeneous OS (real-time and general-purpose) environments \cite{heiser2011virtualizing}, it poses serious challenges, as described in the following.

The development of mixed-criticality systems in the industry has to satisfy stringent requirements provided by safety-critical standards, such as those for the avionics \cite{do178b}, automotive \cite{iso26262}, and railway \cite{cenelec201150128}. These standards refer to temporal and spatial isolation among software components, which are the most critical properties that these systems have to verify. Temporal isolation is about limiting the impact of resource consumption  (e.g., tasks running on a virtual machine) on the performance of other software components (e.g., tasks running on the other virtual machines). Spatial isolation includes the capability of isolating code and data between virtual machines preventing tasks to alter private data belonging to other tasks, including the allocated (memory-mapped) devices. Usually, the above-mentioned standards recommend providing documentation about evidence of a \textit{fail-safe} and/or \textit{fail-stop} behavior for such systems, which ultimately prevent failures leading to human and cost losses.


To face these issues, many solutions and initiatives have been developed over the years, both from industry and academia. This resulted in a variety of, and often partial, solutions that make very hard the decision from industry practitioners to choose a proper virtualization platform, given the domain constraints.  
The main factors that come into play during the process of choosing the proper virtualization solution are the following: hypervisor footprint, which is crucial especially for embedded applications; the compliance with industry safety-related standards; the license of the solution (e.g., proprietary or open-source); the explicit support to high availability; fault tolerance and security; and the supported hardware platforms. 

In the light of the above factors of virtualization for industrial needs, this work surveys the state-of-the-practice of the most representative virtualization approaches adopted or promising for industrial mixed-criticality systems. We group solutions into four main categories:

\begin{itemize}

    \item Solutions based on \textit{separation kernel} and \textit{microkernel}, specifically designed for industrial and embedded domains;
    
    \item Solutions that try to \textit{enhance general-purpose hypervisors} (e.g., Xen and KVM) to support real-time properties, to foster the adoption of mainstream cloud virtualization solutions in the industry.

    \item Solutions that take advantage of the isolation support provided by \textit{Security CPU hardware extensions} (e.g., ARM TrustZone, Intel SGX), to achieve stricter isolation guarantees thanks to the latest hardware extensions;

    \item Solutions based on \textit{lightweight virtualization}, such as \textit{containers} or \textit{unikernels}, which try to achieve a compromise between isolation and the small footprint required in some industrial domains.
    
\end{itemize}

Although there exist other surveys in the current literature, which cover the most common virtualization technologies in embedded real-time domain  \cite{garcia2014challenges, gu2012state, burns2018survey, taccari2014embedded, reghenzani2019real, Struhar5796}, we aim at analyzing existing virtualization platforms and approaches in a different light, considering the common issues that arise in the industry. As already said, examples are isolation properties, real-time performance, testing, and certification issues. \emph{The ultimate aim is to support industrial practitioners to choose the most suitable virtualization solution, according to their specific needs or domain.}

The paper is structured as follows. Section~\ref{sec:background} presents the concept and terms about virtualization, the technical issues for industry application, and discusses related surveys. Section~\ref{sec:dimensions} delineates industrial dimensions with respect to the virtualization paradigm. Section~\ref{sec:solutions} surveys and compares the state-of-the-practice solutions among real-time hypervisors in the light of industrial dimensions. A discussion according to the analyzed solutions, by highlighting the current industrial and scientific trends in virtualization, is provided in Section \ref{sec:discussion}. Section~\ref{sec:conclusion} concludes the paper.

\section{Virtualization in Critical Systems}
\label{sec:background}

Virtualization is among the most promising architectural approaches to implement mixed-criticality systems,  i.e., to integrate software components with different levels of criticality on a shared hardware platform \cite{burns2018survey}. This objective can be achieved using different approaches, such as hypervisors or OS-level virtualization.  
A hypervisor (or Virtual Machine Monitor - VMM) is a software layer that abstracts the hardware resources with the aim to run different and isolated application environments, called Virtual Machines (VMs) or guests, on the same physical machine. A Virtual Machine is an execution environment typically containing an Operating System (OS), called \emph{guest} OS, and the application software.  

\vspace{2mm}

\noindent
$\blacksquare$ \textbf{Taxonomy and applications}. Virtualization approaches can be classified along with several directions. A first distinction is based on the presence of a host OS between the hypervisor and the hardware. A \textit{type-1} hypervisor, often referred to as a ``bare metal'' hypervisor, is run directly on the hardware, acting as a classic OS and controlling directly the hardware resources. A \textit{type-2} hypervisor is executed instead on top of an existing ``host'' OS, which is used to manage hardware resources.

A second distinction is between \textit{full-virtualization} or \textit{paravirtualization}. A fully virtualized hypervisor abstracts completely the hardware resources (e.g., CPU, memory, etc.) to the guests, emulating privileged instructions and I/O operations. Examples are VMware ESXi \cite{vmware_esxi}, KVM \cite{kivity07kvm}, and Microsoft Hyper-V \cite{microsoft_hyperv}. This type of hypervisor has the advantage to let guest OSes or applications run unmodified, as they were running on the physical machine. With paravirtualization, the guest is instead \textit{aware} of the hypervisor. In this case, a guest OS has to be modified to communicate directly with the underlying hypervisor, through so-called \textit{hypercalls}. Such an approach is adopted, for instance, by the Xen hypervisor \cite{XEN_Barham2003}. Similarly to operating systems, hypervisors can be classified as \textit{embedded}, if targeted for a specific application, system, or mission, otherwise they are 
\textit{general-purpose} \cite{heiser2008role}. 

Comparing traditional, \revision{\textit{non--real-time}} hypervisors with \textit{real-time hypervisors}, the latter add explicit support (e.g., specific scheduling algorithms) for the management of the time budged allotted to VMs, in order to assure that individual VMs can comply to stringent and explicit timing constraints. Real-time hypervisors can be further classified as \emph{dynamic} or \emph{static}. Dynamic hypervisors map VM-related resources, like virtual CPUs and virtual memory areas, at run-time as needed. On the contrary, static hypervisors can be seen as configuration layers that partition hardware resources, with a one-to-one mapping between virtual CPUs and physical CPUs, and devices mapped directly into the guest memory areas. Static solutions are often employed for embedded safety-critical or mixed-criticality applications as they are more robust to failures due to misconfigurations and usually introduce less overhead due to virtualization. Moreover, they are usually smaller in terms of lines of code, and thus they are easier to test and certify according to industrial standards.

\begin{figure}[htb]
    \begin{center}
        \includegraphics[scale=0.5]{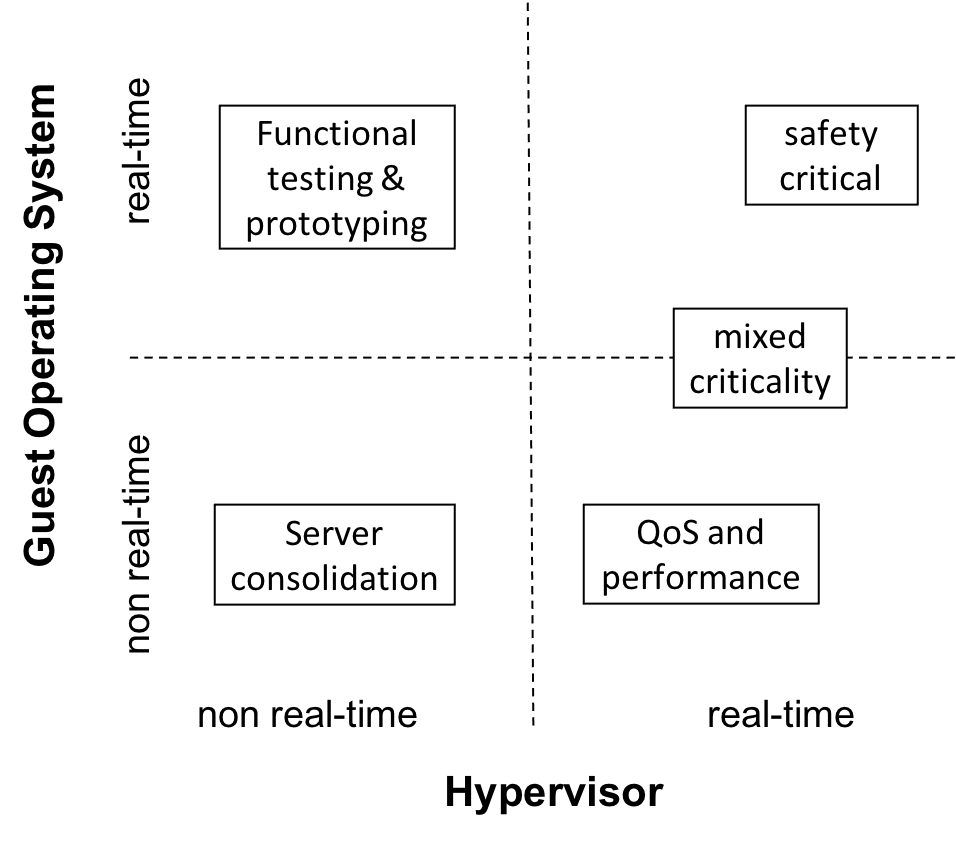}
        \caption{Hypervisor and OS combinations with related applications}
        \label{fig:guests}
    \end{center}
\end{figure}

Different types of guests can be run on top of a hypervisor. Typically, either the guest VM contains a full-fledged OS, or the application software directly \revision{without any need of a VM}, the latter being a more convenient choice in embedded environments. The OS can be either a full-fledged GPOS (General-Purpose OS), such as Linux or Windows, or an RTOS (Real-Time OS). This opens up to different combinations, depending on applications' requirements and constraints (see \figurename{}~\ref{fig:guests}): RTOS and/or GPOS on top of a real-time hypervisor, and RTOS and/or GPOS on top of a non-real-time hypervisor. 

Concerning \figurename{}~\ref{fig:guests}, the focus of this survey is on solutions for the upper-right quadrant, including mixed-criticality systems, being them of major interest for industrial systems.

The lower-left quadrant includes traditional hypervisor application domains, such as server consolidation in cloud environments. Using instead a GPOS on a real-time hypervisor (lower-right quadrant) might be needed to isolate RTOSes running on the same platform, while providing non-critical services in the GPOS, and it might be beneficial if the guest OS has stringent QoS and performance requirements that can be guaranteed with the time budged allocated by the real-time hypervisor. Finally, using an RTOS on a non-real-time hypervisor (upper-left quadrant) is not a common industrial practice, but it might be useful for functional testing, debugging, and prototyping purposes.

A different option that is again gaining popularity lately is to run guests within \textit{unikernels}, which are a promising lightweight solution for embedded domains. \revision{This model does not preclude the selection of a hypervisor since it includes that an} application is linked directly with the OS, treated as a library containing basic functions such as memory management, scheduling, networking stack, and basic I/O drivers. The guest binary will thus embed both the application and the OS code, and can be run directly on the physical hardware or on top of a hypervisor. 

An alternative or complement to hypervisor-based virtualization is OS-level virtualization. The goal is to obtain a virtual domain, called \textit{container}, with its own virtual CPU and virtual memory as in the traditional processes of an operating system, a virtual filesystem, a virtual network, process and user management. These virtual resources are distinct for each \textit{container} in the system.

A container is not a virtual machine in the traditional sense, since there is no emulation of the physical hardware. For this reason, compared to full- and paravirtualization, this type of virtualization is lighter. For this reason, containers are more and more used in cloud environments to \revision{further} improve application consolidation on the same hardware, avoiding replicating the OS stack. For the same reason, container-based virtualization is gaining momentum also in real-time systems \cite{Struhar5796, runx}, especially when stringent scalability and size constraints must be met, providing additional isolation level, while leveraging container orchestration capabilities (e.g., Kubernetes \cite{kubernetes_frakti}). 

\revision{We remark that both container-and unikernel-based virtualization do not include virtualization in the strict sense. However, both containers and unikernels are very spread concepts in the context of virtualization systems literature, and are starting to gain attention in industrial and real-time systems as well, thus we discuss these kinds of approaches in this survey.}

\begin{figure}[htb]
    \begin{center}
        \includegraphics[width=0.45\textwidth]{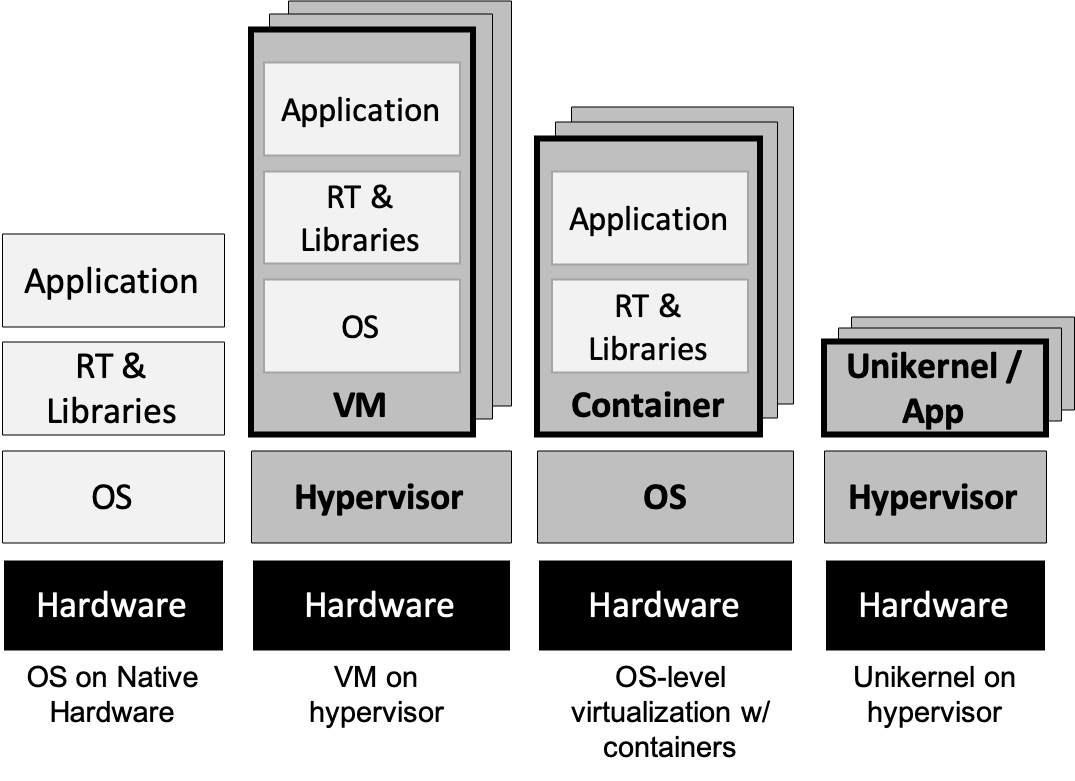}
        \caption{Examples of virtualization approaches.}
        \label{fig:virtualization-approaches}
    \end{center}
\end{figure}

Figure \ref{fig:virtualization-approaches} wraps up the different virtualization approaches discussed above, which are only a partial view of the entire virtualization spectrum.

\vspace{2mm}

\noindent
$\blacksquare$  \textbf{Isolation properties}. 
As mentioned previously, virtualization is one of the enablers for mixed-criticality systems, where in general there is the need to create strongly \emph{isolated partitions} that run applications a different level of criticality. 

In this respect, virtualization must ensure isolation between virtual instances \cite{bugnion2012bringing, inject_hw_fault_hypervisor, jailhouse_perf_isolation_test}. In simple terms, this means that applications running on a virtual domain must have the illusion of being the only ones running on the physical machine. 
In the context of virtualization, we mainly consider three isolation properties.

\emph{Temporal isolation}, or \textit{temporal segregation}, is the ability to isolate or limit the impact of resource consumption (e.g. CPU, network, disk) of a virtual domain on the performance degradation of other virtual domains. This means that a critical task running on a virtual domain (for example a task on a VM or inside a container) must not cause serious delays to other critical and non-critical tasks running in a different virtual domain, avoiding phenomena such as starvation, reduced throughput and increased latency. Temporal isolation is crucial in mixed-criticality systems, where tasks run in a critical domain must guarantee specific performance Service Level Agreements (SLAs) and must not interfere with each other. In the context of safety-critical applications, some standards (e.g., IEC 61508-3 annex F \cite{iec61508}, ISO 26262-6 annex-D \cite{iso26262}, ARINC-653 \cite{arinc653}, DO 178 6.3.3f \cite{do178b}, CAST-32A \cite{CAST-32A}) suggest adopting cyclic scheduling between virtual domains, to assure static and predetermined time slots to each domain.

The other crucial property is \emph{spatial isolation} (also known as \textit{memory isolation} or \textit{spatial segregation}). This property describes the ability to isolate code and data between virtual domains and between virtual domains and hosts. This means that a task should not be able to alter private data belonging to other tasks, including devices assigned to a specific task.
Spatial isolation is usually implemented using hardware memory protection mechanisms, such as the Memory Management Unit (MMU). Considering the case of shared physical devices, also \emph{I/O isolation} becomes important. Often, the IOMMU is used to properly resolve the isolation of memory-mapped devices. In some cases, access to hardware devices from the different virtual domains is serialized.

Finally, \emph{fault isolation}, or \textit{fault/error containment}, prevents that failures, occurring in a virtual domain, are propagated to the hypervisor and/or to other virtual domains, causing blockages or even stopping the whole system.

\vspace{2mm}


\noindent
$\blacksquare$ \textbf{Related surveys}. 
In years, several efforts on real-time virtualization were done, both from the commercial and academic sides. These studies tried to reuse IT virtualization technologies, well tested in cloud computing, for real-time purposes \cite{garcia2014challenges}.
\textit{Gu and Zhao} \cite{gu2012state} survey virtualization technologies for real-time embedded systems including those for safety-critical applications, and discusses technical problems such as task-grain scheduling and lock-holder preemption. \textit{Burns and Davis} \cite{burns2018survey} surveys the state of the art in the field of mixed-criticality systems, with a focus on scheduling problems and solutions for both single- and multiprocessor systems. \textit{Taccari et al.} \cite{taccari2014embedded} discuss embedded real-time virtualization solutions with a focus on ARM hardware-based virtualization support but limit the analysis to open-source projects. \textit{Reghenzani et al.} \cite{reghenzani2019real} present a comprehensive survey on the real-time Linux kernel research (i.e., PREEMPT\_RT). \textit{Struhàr et al.} \cite{Struhar5796} present a technology survey on real-time Linux container technologies, showing the gaps that should be filled to be a viable solution for industrial applications.

In this paper, we analyze the state of the practice of real-time virtualization approaches in the light of industrial needs that include safety/security certification and testing, reuse of legacy systems, and dependability support. To the best of our knowledge, this is often a neglected point of view on the existing solutions' portfolio, which could aid industrial practitioners to choose the most suitable virtualization approach according to their domain requirements and constraints.

\section{Virtualization Dimensions for Industry Needs}
\label{sec:dimensions}



The main question we address in this paper is: \textit{what should be the primary focus of industry when creating a new product or migrating seamlessly legacy systems exploiting virtualization technologies?} In this section, we delineate three main dimensions that industry and researchers should focus on to properly adopt virtualization technology in mixed-criticality real-time domains.


\vspace{2mm}

\noindent
$\blacksquare$ \textbf{Certification \& Testing}.
The development of safety-critical systems raises various challenges from the certification point of view. In order to provide a specific safety integrity level (SIL), 
almost all standards recommend cumbersome V\&V activities. Specifically, several studies in the literature \cite{cotroneo2013fault, cotroneo2012experimental, cotroneo2018run, cotroneo2016faultprog, nfv_dep_guidelines, winter2015no, mazzeo2018sil2} and various international safety-related standards \cite{iso26262, nasastd2004, do178b, ISO/IEC25045, cenelec201150128} provide guidelines for testing activities, which encompass fault injection testing, robustness testing, and performance testing, among other activities, such as error impact analysis, coding standards, code review, etc. 
In a virtualized scenario, such tasks have to be performed at the different levels of the architecture, considering also the use of Commercial-Off-The-Shelf (COTS) components.  
Concerning cybersecurity, some standards require that the final system must satisfy different security requirements in terms of the provided partitioning level, the degree of resource isolation, complete control over communication channels, and the development of auditing mechanisms. This is the case of the Common Criteria for Information Technology Security Evaluation (ISO/IEC 15408) standard \cite{commoncriteria}, which added a profile named “Separation Kernels in Environments Requiring High Robustness” Protection Profile (SKPP) \cite{skpp, zhao2017survey} (that currently was superseded by the NIST document "Security and Privacy Controls for Federal Information Systems and Organizations" \cite{NIST.SP.800-53r4}).

Testing activities like functional and non-functional testing, robustness testing, performance testing, as well as static and dynamic analysis, run-time verification, fault injection, fuzzing, etc., are fundamental during the development of safety-critical systems. Often, these activities are strictly linked to the certification process, but they are usually performed regardless of the objective of making software certifiable. In particular, in real-time virtualization solutions, great relevance assumes the measure of the overhead introduced by the hypervisor, and how it impacts task execution. For example, the worst-case execution time analysis (WCET) could be invalidated due to the newly added software layers, with the consequent need to repeat the analysis. Furthermore, low-level synchronization primitives like spinlocks, and mutexes, could be redefined to prevent problems like priority inversion \cite{prio_inversion} and lock holder preemption \cite{lockholder_problem} problems. These changes might require the original test suite to be reviewed.

In summary, migrating towards a virtualized environment requires redesigning the approaches used traditionally in the industry. The use of certified hypervisors or the availability of test suites may help to reduce the related burden.


\vspace{2mm}

\noindent
$\blacksquare$ \textbf{Reuse of legacy systems}. Legacy systems migration to a virtualization paradigm brings several benefits, such as, avoiding “divorce” of application and legacy OS, allowing the transparent execution of single-core software stacks on multicore hosts, and emulating discontinued hardware. However, the migration requires addressing a twofold issue. First, the pre-existent kernel (GPOS or RTOS) needs to be ported to be properly run on the specific hypervisor. Second, the chosen hypervisor has to support (or needs to be adapted to) the target (or emulated) hardware platform.
These porting issues strictly depend on the type of chosen hypervisor. Indeed, in the case of full-virtualization with complete device emulation, software emulation is needed for each device within the target board; otherwise, the guest OS could not have all the functionalities properly set up. The good point, in this case, is that the pre-existent kernel does not need to be modified. \revision{Paravirtualized solutions are even more affected by this issue since guest OS kernels must be modified according to the interfaces provided with \textit{hypercall}s to handle privileged instructions and virtualize I/O devices.} \revision{Porting issues is also a problem if we consider the unikernel model. Indeed, unikernel image embeds the application and its dependencies. From one side, this reduces traditional compatibility issues between in-VM components, but potentially introduces new issues between unikernel VMs and the target hypervisor and their execution environment (i.e. computing resources, storage and networking prerequisites). Anyway, several hypervisors can be supported if the unikernel images are configured accordingly.}


\vspace{2mm}

\noindent
$\blacksquare$ \textbf{Dependability support}. Virtualization solutions used in cloud computing environments are usually exploited to provide dependable services, such as tenant isolation, high-availability, fault-tolerance, migration and recovery techniques, and, in the worst-case (with severe faulty conditions), graceful degradation of the provided service. Naturally, these features are desirable also for safety-critical real-time systems. Thus, the chosen virtualization solution should support, for instance, \revision{easy-to-use} fault-tolerance tools and redundancy schemes like Triple Modular Redundancy (TMR) or 2-out-of-2, \revision{which are classical schemas used in industry to improve fault-tolerance}.
Concerning security, the hypervisor (and guest OSes) are likely to be the subject of attacks exploiting their vulnerabilities. In general, the addition of new layers may increase the attack surface. This, in turn, can severely harm the safety of the system. Special care must be taken to mitigate attacks, by accompanying the virtualization solution with mechanisms to keep communication channels secure, to cryptographically sign the running code, and provide auditing services.



\section{Virtualization Approaches and Solutions}
\label{sec:solutions}

In this section, we survey representative examples of virtualization solutions, and related approaches, in light of current industry trends and dimensions, mentioned above. This overview is not intended to provide an exhaustive list of current solutions adopted in the industry. Instead, it guides the reader through examples of impactful solutions, in terms of approach and disruption potential, provided not only by software manufacturers through commercial products, but also by research initiatives and experiments with open sources.
Our goal is to highlight the details of each solution to shed light on what is important to know to choose the right approach for given innovation needs.

The examples have been selected among four categories, which resemble four trends in the current or prospective adoption of virtualization in industry domains:

\begin{enumerate}
    
    \item Solutions specifically designed for industrial and embedded domains, that happen to be based on \textit{separation kernel} and \textit{microkernel} approaches;
    
    \item Solutions that try to exploit the best of existing \textit{general-purpose hypervisors}, adapting them to industry needs;
    
    \item Solutions that take advantage of \textit{latest isolation features at the hardware level} (e.g., ARM TrustZone), to achieve the strict isolation guarantees needed by industry standards;
    
    \item Solutions that try to reduce the footprint and save flexibility, with respect to classical virtualization methods, employing \textit{lightweight virtualization} like \textit{containers} or \textit{unikernels}.
    
\end{enumerate}



\subsection{Separation Kernels and Microkernels}




Several solutions have been developed that try to keep down the complexity of the hypervisor while having a strong level of isolation by using virtualization concepts. Developers provide also ad-hoc solutions that apply hybrid virtualization approaches. Such solutions fit well for the embedded domain from dependability, certification, and testing point of view, as well as, they support several board platforms \revision{in order to provide host-level virtualization capabilities by running several guest OSes}. The aim of reduced complexity is usually achieved through the adoption of separation kernel or microkernel approaches.

A \textbf{separation kernel} is a special type of very small bare-metal hypervisor that utilizes hardware virtualization features to \textit{(i)} define fixed virtual machines and \textit{(ii)} control information flows. Separation kernels contain no device drivers, no user model, no shell access, and no dynamic memory; these tasks are all pushed up into guest software running in the VMs. This simple architecture results in a minimal implementation that, while less convenient for desktop or server use, is an excellent fit for embedded real-time and safety-critical systems.
Separation kernels were the first technologies used in years for implementing so-called \textit{partitioned} systems, and various effort was made especially in the avionic domain to support the safety of flight in Integrated Modular Avionics (IMA) according to the ARINC653 standard \cite{arinc653}. Furthermore, the Multiple Independent Levels of Security (MILS) \cite{alves2006mils} provided a high-assurance security architecture intended to allow mixed security applications to be hosted on common hardware. Recent hypervisors bring together separation kernel and virtualization concepts to isolate virtual machines (also named partitions) at different criticality levels on the same hardware platform.

A \textbf{microkernel} is a minimal kernel that implements only a few abstractions and operations that need to be executed in supervisor mode (e.g., memory management, process/thread management, IPC) while handling in user-mode the other kernel functionalities (device drivers, filesystem, networking, paging, etc.). In the last years, several microkernels have been developed for heterogeneous domains, and recently they have been used as hypervisors in order to provide simple partitioning functionalities with increasing reliability and security, using no 3rd-party code and running drivers within guests.

Representative examples of both models are presented in the following.

\vspace{2mm}

$\blacksquare$ \textbf{VxWorks MILS}. VxWorks MILS \cite{vxworks_mils} is a commercial separation kernel provided by WindRiver. The Wind River VxWorks MILS complies with security requirements derived from the SKPP \cite{skpp, zhao2017survey}, and after the SKPP was sunset, it conforms to a subset of SKPP assurance requirements that apply directly to the product that is provided by Wind River to its customers. 
VxWorks MILS supports information flow control, resource isolation, trusted initialization, trusted delivery, trusted recovery, and audit capabilities. The information flow policies are \revision{set} by the customer-defined configuration vector, which includes virtual boards and images, communication channels between virtual boards (direction, mode, etc.), schedules for the virtual boards, authorized system calls for every virtual board, etc.
The MILS establishes two primary security domains – \textit{MILS kernel} (supervisor) space and \textit{virtual board} (user) space. The MILS enforces a \textit{Least Privilege Abstraction Partitioned Information Flow Policy} (PIFP) to ensure security domains access only the resources that are required for their assigned functionality. Allowed information flows between specific virtual boards and resources are specified by the configuration vector and these flows are static. 
The MILS assumes protection from interference and tampering. Indeed, the MILS bootloader (MILS Payload BootLoader) is the \textit{root of trust} of the entire system, and if the validation is successful the MILS kernel (MK) initialization is called; after initialization, the scheduler removes the MK init code and the related part in the configuration vector and schedules the first VB for execution.
At run-time, the VxWorks MILS reference monitor and self-test subsystem verify that the MILS remains in a secure state. If a failure causing the state to become insecure is detected, the MK recovery is invoked to take action as specified by the configuration vector, which can result in rebooting or halting the system. \revision{In \cite{cotroneo2021timing}, \textit{Cotroneo et al.} targeted VxWorks MILS to perform an experimental analysis to establish potential timing covert channels in order to assess the robustness of configurations provided by system designers.}
\revision{In {\cite{aroca2009real}}, \textit{Aroca et al.} assess VxWorks MILS realtimeliness by overloading the MILS kernel with $\sim$ 400 tasks alongside a ping flood was executed against a testbed that uses a signal generator and analyzes the signal response with an oscilloscope. The results show that the testbed could reliably handle and measure a $260KHz$ input frequency, with the worst response time of $3,85$ $\mu$s.}

\vspace{2mm}

\vspace{2mm}

$\blacksquare$ \textbf{PikeOS}. PikeOS \cite{pikeos} is a commercial hypervisor from SYSGO, used in the avionic domain. PikeOS architecture is based on the L4 microkernel and can run on Intel x86, ARM, PowerPC, SPARC v8/LEON, or MIPS processors. PikeOS supports multi-core platforms natively. That solution adopts three different kinds of scheduling algorithms, that is, priority-based, time-driven, and proportional share. For each real-time VM (critical VM), PikeOS statically assigns a time-slice; whenever a critical VM does not have any task to be executed, it donates CPU time to non-critical VMs. PikeOS architecture is ARINC653-compliant, in the sense that the PikeOS microkernel is the only privileged software and it is in full control of the virtual partitions. PikeOS supports several guest OSes like Android, RT-POSIX, ARINC653-based, Java, RTEMS.
PikeOS has been the target and the basis of several academic and industrial evaluations. For example, \textit{August} \cite{august2014idp} analyzes the effect of a cache-timing side channel attacks on AES \cite{heron2009advanced} focusing on a virtualization scenario based on PikeOS, as an example of a real-time system dedicated to security. It will furthermore evaluate methods to counteract that threat by using the system’s scheduler. Regarding certification and formal verification, in \cite{verbeek2015formal} the authors formalized the hardware-independent security-relevant part of PikeOS in order to prove intransitive noninterference properties \cite{roscoe1999intransitive}; moreover, in  \cite{baumann2009verifying} the authors presented first results in the verification of the PikeOS microkernel system calls. \revision{In {\cite{muttillo2019benchmarking}}, \textit{Muttillo et al.} leverage Dhrystone benchmark to compare Xtratum and PikeOS by varying the compilation optimization flags (e.g., O0, O1, etc.). We suggest the reader to refer \cite{muttillo2019benchmarking} for the extensive results provided by the study.}




\vspace{2mm}

$\blacksquare$ \textbf{Xtratum}. Xtratum \cite{masmano2009xtratum} is a para-virtualized type-1 partitioning hypervisor very popular for avionic embedded safety-critical systems and consists in $\sim 9K$ LOC. Xtratum is based on the APEX model, defined within the ARINC 653 standard \cite{arinc653}. Furthermore, Xtratum supports several CPUs like Intel x86 family, SPARCv8 family, ARMv7, and ultimately RISC-V that is under development.
Xtratum provides temporal isolation between virtual partitions by leveraging a  fixed cyclic scheduler. Spatial separation is provided by forcing partitions to execute in user-mode without any memory shared area. Xtratum data structures are all pre-defined at build time through a configuration file, in order to know exactly what resources the hypervisor will use. 
Xtratum defines a minimum set of hypercalls each of which has a known execution time. Finally, Xtratum enables interrupts only for partitions currently running, in order to minimize temporal interferences.
Regarding the software certification, Xtratum hypervisor was used as a fundamental component for developing ARINC653-compliant RTOS \cite{zamorano2010open, esquinas2011ork} and for porting an OSEK-VDX-based RTOS to run on top of Xtratum \cite{oversee}. Furthermore, the research community leverage Xtratum as a basis for fault-tolerant platforms in the context of embedded systems for space applications \cite{campagna_xtratum}. \revision{A preliminary analysis on the realtimeliness of Xtratum is conducted in \cite{zhouinitial}, however is not very indicative. Despite, \textit{Carrascosa et al.} \cite{carrascosa2014xtratum} provide experimentation for native versus partitioned applications with the aim of evaluating the performance loss due to the presence of Xtratum hypervisor. The authors compare the execution time of Dhrystone and CoreMark benchmarks on bare-metal and a partition under the Xtratum cyclic scheduling. The results show an execution time of $\sim$ 1-10 seconds, with $0.008$\% and $1.087$\% a performance loss for Dhrystone and CoreMark respectively. Further, the authors evaluated the partition context switch (PCS) impact, and it is estimated to be in the range of 149 to 151 $\mu$s.}

\vspace{2mm}

$\blacksquare$ \textbf{Jailhouse}. Jailhouse \cite{jailhouse} is a Linux-based partitioning hypervisor developed within a research project by Siemens publicly available in 2013. In particular, Jailhouse enables asymmetric multiprocessing (AMP) cooperating with the Linux kernel in order to run bare-metal applications or guest OSes properly configured. Given the Jailhouse objective more related to isolation than virtualization, the hypervisor splits physical resources (CPUs, memory, I/O ports, PCI devices, etc.) into strongly isolated compartments called \textit{cells}. Each \textit{cell} is exclusively assigned to one guest OS and its applications called \textit{inmates}. Jailhouse includes a cell, called \textit{root cell}, that runs the Linux kernel and will execute the Jailhouse hypervisor itself and the other cells. Despite the main objective being partitioning resources, Jailhouse allows inter-cell communication through the \textit{ivshmem} device model \cite{ivshmem} from the QEMU project \cite{qemu}, which is based on an abstraction of PCI devices.

Jailhouse consists of a few lines of code (around 30K \textit{C} and 1K \textit{Assembly} lines of code), thus it should be ease both the process of certification and the application of formal methods for verification. Further, Jailhouse is released applying continuous integration and static code analysis tools. 
Jailhouse supports various OSes besides Linux, like L4 Fiasco.OC on Intel x86 and FreeRTOS on ARM, Erika Enterprise RTOS v3 on ARM64, and several ARM-based boards (e.g., NVIDIA Jetson TK1, Xilinx ZCU102, etc.). 
About fault-tolerance, Jailhouse provides a simple mechanism that allows restarting non-root cells as soon as they enter deadlock states detected using timeouts. Recently, Jailhouse was chosen as the building block for proposing a new family of safety-critical computing platforms designed to be compliant with IEC 61508 standard \cite{selene_project}.

\revision{In \cite{jailhouse_perf_isolation_test}, Jailhouse is the target of real-time assessment. The authors define an isolation coefficient, which represents the resulting slow-down due to the execution of tasks in the presence of other running tasks. For CPU isolation tests, an isolation coefficient of $0.40$ and $0.0086$ are provided respectively by the Linux and Jailhouse. The authors perform also L2-cache contention tests with basically no difference in execution time nor cache misses. 
Finally, the authors provide memory bus isolation tests, and the results show that Jailhouse does not provide any mechanism of bus isolation despite it does not introduce any overhead penalties.}

\vspace{2mm}

%

%

\revision{$\blacksquare$ \textbf{L4-based}}. NOVA \cite{steinberg2010nova} is a type-1 hypervisor written in C++, developed to enhance security more than safety. The NOVA design is very similar to the L4 microkernel, but in contrast, it provides a full-virtualization solution. NOVA splits the hypervisor in a full-privileged critical component named \textit{micro-VMM}, while the rest of the components are not privileged. The micro-VMM includes only the scheduler (in that case NOVA uses a preemptive priority-driven round-robin scheduler with one runqueue per CPU), MMU, a limited set of hypercalls, and implements the communication mechanisms between itself and other non-privileged components. In total, the size of the NOVA hypervisor settles down in 36KLOC including the microhypervisor (9 KLOC), a thin user-level environment (7 KLOC), and the VMM (20 KLOC). 
In general, the NOVA authors discuss deeply how the design principles can prevent several virtualization attacks like VMM attacks, guest attacks, and so on. They analyzed the performance overhead introduced in NOVA and demonstrate that it can be lower than 1\% for memory-bound workloads. \revision{Further, they evaluate NOVA against IPC and virtual TLB miss microbenchmarks, according to different CPU architectures, obtaining latencies in the order of $100ns$.} Finally, they compared memory virtualization using hardware-based nested paging to a shadow page tables approach and observed that nested paging reduces the virtualization overhead from more than 20\% to 1–3\%. NOVA supports many ARMv8-based boards (e.g., 
NXP i.MX 8MQuad, Renesas R-Car M3, Raspberry Pi 4 Model B, Avnet Xilinx Ultra96, as well as QEMU virtual platform) and x86/x68\_64 CPU families.
NOVA was also analyzed by \textit{Tews et al.} in the context of the European project named Robin \cite{tews2008nova} for formal verification purposes. The objective is to develop a semantic compiler in order to provide denotational semantics for C++ which includes all the C++ primitive data types of NOVA hypervisor.

\revision{The seL4 {\cite{klein2009sel4, elphinstone2013l3}} is a formal-verified microkernel designed to be used in security- and safety-critical systems. In particular, sel4 is functionally correct against a formal model enforcing both integrity and confidentiality; timing channels proofs are still under assessment \cite{heiser2019can}. seL4 uses a priority-based scheduling policy and implements \textit{scheduling-context capabilities} for assigning CPU time in the context of mixed-criticality systems \cite{lyons2018scheduling}. In particular, a component can only obtain CPU time if it holds the scheduling-context capability, which specifies also the amount of CPU time that can be used. A scheduling context consists of a time budget (i.e., a time slice) and a time period that determines how often the budget can be used. A thread will not get more time than one budget per period. Further, sel4 leverages both ARM and x86 virtualization extensions to provide interfaces to support running virtual domains, which are implemented in user space. These interfaces compose the VMM, which initializes memory and provides exception handlers for emulated device drivers.  The VMM was recently redesigned from a simply sel4 application to a set of CAmkES (Component Architecture for Micro-Kernel-based Embedded Systems) components {\cite{camkes}}. sel4 is provided with a WCET analysis that results in determinist upper bounds for system calls and interrupt latencies {\cite{blackham2011timing, sewell2017high}}. In particular, \textit{Blackham et al.} {\cite{blackham2011timing}} provided an evaluation of seL4 and obtained that sel4 provides a guaranteed interrupt response time of around 500 $\mu$s on a BeagleBoard-xM platform with an ARM Cortex-A8 core. In open systems (arbitrary code can execute on the system), the interrupt response time is about 2 ms. Finally, different efforts are made to enhance sel4 with fault-tolerance capabilities. In particular, researchers proposed a mechanism to provide both task backup and recovery, as well as two checkpoint-based optimization strategies \cite{luan2018towards, xu2016towards}.}

\vspace{2mm}

\vspace{2mm}
\begin{mybox}{\textbf{Separation Kernel}}

$\tikztriangleright[blue,fill=gray!80]$ Separation kernels are designed to provide high levels of isolation coupled with dependability support. However, these solutions are not meant to be deployed on cloud platforms.

$\tikztriangleright[blue,fill=gray!80]$ The development of separation kernels is often strictly related to the certification process. About testing, some studies provide an evaluation of the overhead due to virtualization, assessment of the isolation and recovery mechanisms.

\end{mybox}

\begin{mybox}{\textbf{Microkernel}}

$\tikztriangleright[blue,fill=gray!80]$ Microkernels used as hypervisors are mainly designed to reduce at the maximum the trusted computing base compared to classical full virtualization solutions, providing high security.

$\tikztriangleright[blue,fill=gray!80]$ Since microkernels are lightweight solutions, they are well suited for formal verification and testing activities, which makes easy the certification process.   

$\tikztriangleright[blue,fill=gray!80]$ 
\revision{Microkernels (like sel4) provides real-time capability with memory protection, for security, as well as part of its support for mixed-criticality systems.}

\end{mybox}





\subsection{General-purpose Hypervisors}


In the last decades, Xen and KVM hypervisors were among the most used solutions in server virtualization. Xen is a type-1 hypervisor that provides paravirtualization technologies. It was the first attempt to overcome the performance penalty due to the dynamic binary translation \cite{XEN_Barham2003}. On the other side, KVM  is one of the most used hardware-assisted virtualization solutions, which exploits hardware extensions provided by modern CPUs. For example, the Intel VT-x enables the CPU to execute in two modes, i.e., the \textit{non-root} mode used to run guest OSes code, and the \textit{root} mode used to run the hypervisor. As soon as a VM attempts to execute privileged instructions (prohibited in non-root mode), CPU switches to root mode in a trap-like way to properly handle the instruction \cite{neiger2006vtx}. KVM is by definition a type-2 hypervisor since it requires the Linux kernel, but in practice, it acts as a type-1 hypervisor since it takes full control of the underlying hardware. It uses QEMU to provide I/O device emulation.

Despite both Xen and KVM are general-purpose hypervisors, they are currently used as tailored and working solutions for embedded systems and real-time clouds, if properly tuned \cite{abeni2020using, abeni2019experimental,real_time_kvm, xi2015rt,agl_kvm}, as described in the following.

\vspace{2mm}

$\blacksquare$ \textbf{Xen}. In Xen, the main approach has been to optimize the scheduling algorithms of the virtual CPUs and to improve the interrupt handling \cite{RT-XEN_xi2011rt, RT-XEN_xi2014real, jeong2011parfait, gupta2006enforcing, govindan2009xen}. By default, Xen adopts the \textit{Credit} scheduler, which is a (weighted) proportional fair share virtual CPU scheduler. The user could tune the CPU share for each domain. Furthermore, the scheduler load balances the workload among vCPUs. RT-Xen \cite{RT-XEN_xi2011rt, RT-XEN_xi2014real} is one of the most important examples of using Xen for real-time purposes by providing a hierarchical real-time scheduling framework for Xen. In \cite{RT-XEN_xi2011rt}, the authors provided an empirical study on fixed-priority hierarchical scheduling in Xen, focusing on four real-time schedulers: Deferrable Server, Periodic Server, Polling Server, and Sporadic Server. They demonstrate that \textit{Deferrable Server} is more suitable for soft real-time applications, while \textit{Periodic Server} is the worst under the overloaded scenario.  
RT-Xen is at $2.2$ version (last update in 2015), supporting both RM and EDF scheduling policy. The developers re-implemented the RM scheduling policy inside the RTDS scheduler in Xen $4.6$ (RTDS is still an in-development feature). This effort is to improve the efficiency of the implementation of the RM scheduling policy and synchronize RT-Xen with the latest Xen version. Further, developers implemented also the \textit{null} scheduler, which makes Xen a partitioning hypervisor, by statically assigning a single vCPU to a specific pCPU, removing any scheduling decision.
Recently, Xen was used as a building block for Xilinx embedded systems \cite{xen_ultrascale}. Xilinx chooses Xen due to several motivations: \textit{(i)} it is a robust and reliable solution; \textit{(ii)} recent developments of Xen takes full advantage of ARMv8 and his virtualization extension (around $30$KLOC for specific hardware configuration), as well as all the support for the ARM System Memory Management Unit (SMMU); \textit{(iii)} it is provided with a free-of-use license and has an active user and developer community. 
In years, Xen developed the \textit{Xen Test Framework (XTF)} \cite{xen_test_framework}, a framework for both creating microkernel-based tests and a suite of tests built using the framework itself: pre-built tests include assessment of specific security vulnerability, sanity checks, and functional tests. Further, Xen developed also CI platform called \textit{OSSTest} \cite{osstest_xen}, to run automatically test cases and leverage CI tools. 
Finally, Xen brings various efforts for safety certification aspects, such as the DornerWorks Xen-based hypervisor named \textit{ARLX}, which is ARINC653 compliant \cite{vanderleest2013safe}. Recently, the Xen FuSa Special Interest Group (\textit{FuSa SIG}), which includes the Xen Project community together with industry vendors and safety assessors, provided objectives and high-level agreements to build and certify safety-critical systems (mainly in the automotive domain) based on mainline Xen hypervisor codebase \cite{xen_fusa_sig}. \revision{In \cite{abeni2019experimental}, \textit{Abeni et al.} run the \textit{cyclictest} as stress load in scenarios with non--real-time and real-time kernels used at guest and Dom0 level. In particular, they used the default Xen scheduler and assign dedicated pCPUs to the DomUs. The results show that using Xen’s HVM virtualization mechanism can result in very high latencies in presence of some load in Dom0 (in the order of seconds), leading to unusability of Xen in the real-time domain. However, this issue can be avoided by using PV or PVH modes. Indeed, Xen allows reaching latencies in the order of $100$ and $200$ $\mu$s for PV and PVH modes respectively.}



\vspace{2mm}


$\blacksquare$ \textbf{KVM}. KVM-based solutions are mainly based on patching the host Linux kernel or improving KVM itself in order to comply with real-time constraints. 
The {\lmttfont PREEMPT\_RT} \cite{preempt_rt} is a set of patches of the Linux kernel, which provide real-time guarantees (e.g., predictability, low latencies) still using a single-kernel approach, against co-kernel model \cite{reghenzani2019real}. \revision{The main idea behind the co-kernel approach is to have another OS working as a layer between the hardware and the GPOS kernel, which intercepts interrupts and route them to real-time tasks or to GPOS tasks. Then, the scheduler must guarantee that real-time tasks do not miss deadlines.} Instead, the {\lmttfont PREEMPT\_RT} patch provides several mechanisms like high-resolution timers, threaded interrupt handlers, priority inheritance implementation, Preemptible Read-Copy-Update (RCU), real-time schedulers, and a memory allocator.

\textit{Kiszka et al.} \cite{kiszka2009towards} developed a para-virtualized scheduler at the task level, which allows the scheduler to cooperate with KVM via two new hypercalls, in order to manage threads at different priorities. They use KVM as a real-time hypervisor by assigning higher priorities to real-time threads within a VM, while lower priorities to threads running at the host layer. \textit{Cucinotta et al.} \cite{cucinotta2009respecting} developed a scheduling algorithm by extending the Linux cgroups interface \cite{linux_cgroups}. The authors proposed a variant of the CBS (Constant Bandwidth Server)/EDF scheduler to be used for inter-VM scheduling (at hypervisor level), and a fixed-priority scheduler within each VMs. In \cite{cucinotta2010providing}, the same authors focused on I/O issues. The idea was to group in the same reservation both VM threads and KVM threads and kernel threads needed for I/O virtualization (e.g., network or disk). \textit{Zhang et al.} \cite{zhang2010performance} applied various real-time tuning to Linux host by using the {\lmttfont PREEMPT\_RT} patch. They focused on a dual-guest scenario, in which they consolidated an RTOS and GPOS on the single KVM instance.
Recently, KVM was supported in automotive industrial scenarios by the Automotive Grade Linux (AGL), which is a collaborative open-source project to accelerate the development of the connected car \cite{agl_kvm}.
Regardless of all the solutions based upon {\lmttfont PREEMPT\_RT}, currently, this patch is accompanied by several test cases provided by the Linux Test Project (LTP) \cite{ltp_rt_tests}, and benchmarks about, among others, worst-case latency scenarios, latency debugging with tracing, approximating application performance, scheduling attributes tests, and tests against classic three-way priority inversion deadlock. \revision{In \cite{abeni2019experimental}, the authors analyzes the ability of KVM to serve real-time workloads. The results show that KVM causes worst-case latencies smaller than $100$ $\mu$s. In general, the authors suggest using real-time kernels both at guest and host level.}

\vspace{2mm}

\begin{mybox}{\textbf{General-purpose Hypervisors}}

$\tikztriangleright[blue,fill=gray!80]$ Could be adapted for real-time purposes through patches and re-design of specific critical components like CPU emulation.

$\tikztriangleright[blue,fill=gray!80]$ They are a good choice when there are requirements related to cloud computing, like VM migration, orchestration, and high-availability mechanisms. 

$\tikztriangleright[blue,fill=gray!80]$ KVM- and Xen-based hypervisor solutions should be carefully tuned to prevent higher latencies introduced by the scheduling approach and emulation (CPU and I/O) mechanisms. 

$\tikztriangleright[blue,fill=gray!80]$ Explicit support for testing is available, and certification aspects have been started to be a primary focus, especially in Xen.

\end{mybox}

\subsection{ARM TrustZone-assisted Virtualization}

In order to increase the isolation of virtual domains, the research community explored the possibility of leveraging \emph{hardware-assisted} solutions for security purposes in the safety-critical domain. ARM with TrustZone \cite{trustzone} and Intel with SGX \cite{intel_sgx} provide the most used architectures. Normally, these solutions enable a so-called Trusted Execution Environment (TEE) and provide confidentiality and integrity. 

In literature, there were very few studies that leverage Intel SGX extensions to design real-time mixed-criticality systems. One study that is worth mentioning is provided with a positioning paper by \textit{De Simone et al.} \cite{de2019isolating}, which explored the possibility of using the SGX to enforce the isolation among critical tasks running on top of unikernel-based hypervisor \cite{madhavapeddy2013unikernels, madhavapeddy2013unikernels_2}.
The most explored approach has been to use the security features of ARM TrustZone. This technology supports two virtual execution states (i.e. ``secure'' and ``non-secure'') and provides time and spatial isolation between the two environments  \cite{douglas2010thin, frenzel2010arm,  pinto2017ltzvisor, oh2012acceleration, schwarz2014affordable}. In particular, for virtualization purposes, the \textit{non-secure world} and the \textit{secure world} are used for running different VMs that are managed by the hypervisor software that runs in the \textit{monitor mode}. Mostly, researchers used ARM TrustZone with a \textit{dual-guest OS} configuration for running side by side a general-purpose OS (GPOS) within the non-secure world and a real-time OS (RTOS) in the secure-world having higher privileges. This way, critical tasks running on top of the RTOS are isolated from non-critical tasks. 

\vspace{2mm}

$\blacksquare$ \textbf{LTZvisor/RTZvisor}. One of the most representative TrustZone-assisted virtualization solutions is LTZvisor that is designed mainly for mixed-criticality systems \cite{pinto2017ltzvisor}. LTZVisor implements the dual-guest OS scenario, in which the RTOS and GPOS share the same physical processor, but the GPOS is scheduled only when RTOS is idle. An improved version of LTZVisor \cite{pinto2017lightweight} supports asymmetric multi-processing execution, in which the RTOS and hypervisor execute in one core within the secure world, while another core runs the GPOS within a non-secure world. In that case, the authors avoid starvation of GPOS tasks.
The first version of LTZVisor consists of less than 3KB of memory footprint and introduces a GPOS performance degradation of around 2\% for a 1-millisecond guest-switching rate. \revision{In \cite{pinto2017ltzvisor}, the authors evaluate several latency-sensite operations. In particular, \textit{i)} partition-switch operations take $\sim$ 20 $\mu$s, assuming no real-time tasks ready to run once the RTOS is rescheduled; \textit{ii)} the process of checking that no real-time tasks are ready to run and then trigger the switch to the non-secure world takes $\sim$ 12 $\mu$s; \textit{iii)} switching from the RTOS to the GPOS takes $\sim$ 3 $\mu$s; \textit{iv)} the hypervisor guarantees $\sim$ 2 $\mu$s of interrupt latency in the case of serving FIQs (Fast Interrupts, which are serviced first when multiple interrupts occur) while GPOS is running, and a total of $\sim$ 5 $\mu$s to restore RTOS execution.}

The same authors proposed RTZVisor \cite{pinto2016towards} and his successor $\mu$RTZVisor \cite{martins2017murtzvisor} as a solution for multi-guest OS scenario. In that case, the hypervisor software still runs in the monitor mode, while each of the guest OSes can run switching between the non-secure and secure worlds. Specifically, the active guest OS runs in the normal world, while the context of inactive guests is preserved in the secure world. \revision{$\mu$RTZVisor supports both coarse-grained partitions that run guest OSes on the non-secure world, and user-level finer-grained partitions on the secure side that are used for executing \textit{secure tasks} implementing kernel extensions. The adopted scheduler is based on \textit{time domains}, which are execution windows with a constant and guaranteed bandwidth. At each time domain is assigned an execution budget and each domain is scheduled according to round-robin policy. Further, the scheduler allows assigning partitions to the \textit{domain-0} time window, in which partitions are scheduled in a priority-based, time-sliced manner. The \textit{domain-0} can preempt partitions running in different domains. In \cite{martins2017murtzvisor}, \textit{Martins et al.} evaluate time switching between guest partitions and secure tasks or between secure tasks; the results show that the switching process takes about $19.4$ and $10.4$ $\mu$s, respectively. $\mu$RTZVisor provides in the worst case about $180$ $\mu$s of interrupt latencies. Finally, about IPC the authors evaluate both asynchronous and synchronous communication. In particular, they analyzed the time the running partition needs to perform the Send, Receive and SendReceive hypercalls from a guest partition. Considering a 64-byte message size, the hypercall execution time is in the order of $5$ $\mu$s for each operation.}

\vspace{2mm}

$\blacksquare$ \textbf{VOSYSMonitor}. VOSYSmonitor \cite{Lucas2018vosys, lucas2017vosysmonitor} is a low-level closed-source software layer that executes in the monitor mode of the ARM TrustZone architecture. It was conceived for the automotive industry and it is compliant with the ASIL-C requirements of the ISO 26262 standard \cite{iso26262}. VOSYSmonitor enforces the RTOS, or safety-critical OS, to run on the secure world, while multiple non-critical guests can run on the normal world, managed by a non-real-time hypervisor (e.g., Xen or KVM). Non-critical guests can run only when the critical OS releases the permission to run on the assigned core in the normal mode. Context switches are efficiently managed, through interrupt handling, in the monitor mode. 
To achieve the required level of certification, VOSYSmonitor implements several \textit{safety features}. Among them, we mention mechanisms for safe core synchronization, runtime self-tests (e.g., to check memory and I/O isolation properties, code integrity and performance monitoring), and the introduction of a \textit{safe state} which is used to preserve the proper execution of the critical OS in the secure world in case a fault is detected by the runtime self-tests. Among possible measures, the safe state includes the switching off of appliances in the normal world and the migration of the secure world from a core to another. \revision{In \cite{lucas2017vosysmonitor}, the authors evaluate VOSYSMonitor against the ARM Juno R1 and the Renesas R-CarH3 platforms, by analyzing the context switch latency using the ARMv8 Performance Monitoring Unit (PMU). The results for the Juno board shows that VOSYSmonitor is $\sim$ 100\% and $\sim$ 200\% faster than ARM Trusted Firmware (ATF) \cite{arm_trusted_firmware} respectively in VOSYSmonitor running on Cortex-A57 with interrupt handler and almost in VOSYSmonitor running on an A-53 core without interrupt handler, with overall latencies in the order of $0.5$-$1$ $\mu$s. Considering also the context switch including the FIQs, VOSYSMonitor settles around $200 ns$ for interrupt latencies.}

\vspace{2mm}

\begin{mybox}{\textbf{ARM TrustZone-assisted Virtualization}}

$\tikztriangleright[blue,fill=gray!80]$ ARM TrustZone enables virtualization thanks to \textit{dual world} execution model. 

$\tikztriangleright[blue,fill=gray!80]$ TrustZone-based solutions are strictly linked to the specific ARM CPU architecture, thus they are not suitable for supporting other platforms (e.g., PowerPC, Intel). 

$\tikztriangleright[blue,fill=gray!80]$ These solutions are provided with well-defined test suites and performance analyses, as well as approaches for failure recovery.


\end{mybox}

\subsection{Lightweight Virtualization}

In some cases, the stringent footprint requirements of embedded mixed-criticality systems call for a lightweight virtualization approach. For this reason, lightweight solutions based on \textit{OS-level virtualization with containers} and \textit{unikernels} are starting to be explored in industrial domains.

Adopting \textbf{OS-level} or \textbf{container-based} virtualization in real-time domain is a recent trend. The goal is to leverage \textit{containers} in lieu of VMs to achieve isolation with small footprint in mixed-criticality systems \cite{rt-containers2020}. In many cases, it is indeed not necessary to replicate an entire OS within a VM, especially if specific OS functionalities are not needed. The key idea is to enhance the abstractions of OS processes (called \textit{containers}), by extending the (host) OS kernel. For example, Linux leverages the \textit{namespace} process isolation mechanisms \cite{man2018namespaces} and \textit{cgroups} that provides resource management capabilities \cite{linux_cgroups}. A container will have its virtual CPU and virtual memory (like in traditional OS processes), but also virtual filesystem (i.e., the container perceives a filesystem structure that is different than the host’s), virtual network (i.e., the container sees a different set of networking interfaces), IPC, PIDs, and users management. These virtual resources are distinct for each container in the system. The approach is gaining popularity also in the context of consolidated real-time platforms, such as VxWorks by WindRiver, now featuring a container engine compliant with OCI (Open Container Initiative - opencontainers.org).

In the literature, Linux-based real-time container solutions mainly adopt two approaches: \textit{(i)} the use of co-kernels, and \textit{(ii)} the modification of the Linux scheduler.

\vspace{2mm}

$\blacksquare$ \textbf{RT-CASE}. RT-CASE \cite{cinque2019rt} is built using the co-kernel approach. Indeed, the real-time tasks run within real-time containers (named \textit{rt-case}) and will be scheduled by the co-kernel. That approach exploits co-kernels that are known to provide better real-time performance and functionalities, while keeping all the mechanisms and tools provided by a container engine. Each \textit{rt-case} is assigned with a criticality level, and tasks with a lower criticality level must not interfere with tasks with a higher criticality level.
RT-CASE architecture includes container management tools and libraries, and a \textit{feasibility checker} that is responsible for admitting a new container on a compute node according to already running real-time containers.
At the kernel level, RT-CASE leverages the dual-kernel approach, by using a co-kernel like RTAI or Xenomai. The co-kernel makes the host kernel fully preemptable, thus both general-purpose containers and host tasks will be preemptable by real-time tasks and containers.
The \textit{rt-lib} is a key component in RT-CASE, and it provides the mapping between real-time tasks on real-time CPUs according to the container criticality level. Furthermore, \textit{rt-lib} provides standard primitives to run non-modified tasks within real-time containers. Finally, RT-CASE is designed to migrate on-demand real-time containers on nodes within a large-scale cloud platform.

\vspace{2mm}

$\blacksquare$ \textbf{Hierarchical scheduling of containers}. \textit{Abeni et al.} \cite{abeni2019container} proposed the use of real-time containers by modifying the Linux scheduling mechanism to provide two levels of hierarchical scheduling. First level Earliest Deadline First scheduler selects the container to be scheduled on each CPU. Subsequently, the second level Fixed Priority scheduler selects a task in the container. CPU reservation (run-time quota and period) is assigned to each of the containers. \revision{In \cite{abeni2019container}, \textit{Abeni et al.} provide a Real-Time Schedulability Analysis proving that using the proposed hierarchical scheduler the all the tasks running at guest level consume all the runtime assigned to the vCPUs of the VM. Further, they perform an experiment to show the advantages of using the proposed scheduler for the management of a real-time JACK audio processing workflow.}
\textit{Cucinotta et al.} \cite{cucinotta2018virtual, cucinotta2019reducing} leveraged this hierarchical real-time scheduler for providing preliminary results of an ongoing project about using a container-based solution in Network Function Virtualization infrastructures \cite{nfv_bench}. The authors proposed a mechanism to reduce temporal interferences among concurrent real-time services deployed on containers, and evaluated the proposed approach by using LXC containers \cite{lxc}. The results show stable performance of deployed services, enabling the possibility to apply sound performance modeling, analysis, and control techniques. \revision{Also in that case, the authors provide a schedulability analysis using a hierarchical real-time scheduler, which provides predictable QoS and can be used for real-time workloads.}

\vspace{2mm}

\begin{mybox}{\textbf{OS-level Virtualization}}

$\tikztriangleright[blue,fill=gray!80]$ Gained popularity in recent years due to the provision of lightweight isolation solutions in embedded systems, by leveraging the host OS as a hypervisor. 

$\tikztriangleright[blue,fill=gray!80]$ Solutions exploit built-in dependability mechanisms like container migration and load balancing, as well as container recovery (e.g., restart), at the expense of lower security. 

$\tikztriangleright[blue,fill=gray!80]$ Despite the use of CI tools and well-defined test suites, these solutions require more analysis and studies in the industrial context, especially in the view of certification and isolation testing tasks.

\end{mybox}

In order to increase isolation, performance, and security it is possible to run a single application in its virtual domain. Such a model is known as \textbf{unikernel} or \textbf{library OS}, in which the full software stack of a system, including OS components, libraries, language runtime, and applications, are compiled into a single VM that runs directly on a general-purpose hypervisor (e.g., Xen). This approach introduces benefits such as a high performance, small code base, and a reduced certification effort, due to the low amount of software to be verified. However, stronger isolation proofs to be reported to certifiers are still lacking. Further, the attack surface of unikernel instances is small, as they lack the variety of functions provided by standard OSes, as well as the tools used to exploit them (no shells, utilities, etc.).
 
A fundamental drawback of unikernels is that developers must manually port target applications to the underlying minimal OS. This brings significant engineering effort since it takes a considerable amount of time and needs experts with high knowledge of underlying OS details. \revision{\textit{HermiTux} {\cite{olivier2019binary,olivier2021syscall}} is a solution that tries to mitigate porting issues in unikernel-based systems. In particular, HermiTux emulates OS interfaces at runtime accordingly to the Linux ABI, and runs a customized hypervisor-based ELF loader to run a Linux binary side-by-side with a minimal kernel in a single address space VM. All the system calls made by a program are redirected to the implementations the unikernel provides. Hermitux supports multithreading and SMP, as well as, checkpoint/restart and migration, which are crucial for orchestration purposes. Another relevant example for circumventing issues mentioned before is \textit{Unikraft} {\cite{kuenzer2021unikraft}}, which provides} a highly-configurable unikernel code base for speeding-up development.

Despite existing several unikernel solutions most suitable for cloud computing scenarios \cite{mirageos, kivity2014osv, rumprun, wick2012halvm}, \revision{representative examples include \textit{ClickOS} and \textit{HermiCore}.}

\vspace{2mm}

$\blacksquare$ \textbf{ClickOS}. \revision{\textit{ClickOS} {\cite{martins2014clickos}} is an example of using unikernel model in the industry. NEC Ltd. proposed this solutions for consolidating several high-performance virtualized network middleboxes on top of Xen {\cite{martins2014clickos}}. In particular, \textit{ClickOS} is based on MiniOS unikernel {\cite{popuri2014tour}} and} brings a number of optimizations to the Xen’s network I/O sub-system in order to perform fast networking for traditional VMs; in particular, ClickOS includes (i) replacing Open vSwitch back-end switch with a high-speed ClickOS switch, (ii) removing the \textit{netback} {\cite{xen_networking}} driver from the pipe, still used as control plane driver to perform actions such as communicating ring buffer addresses (grants) to the \textit{netfront} driver, and (iii) changing the VM \textit{netfront} driver to map the ring buffers into its memory space. 

\revision{$\blacksquare$ \textbf{HermitCore}. \textit{HermitCore} \cite{lankes2016hermitcore} is a unikernel solution designed for High-performance Computing (HPC) scenarios and particularly for NUMA architectures. This solution leverages a library OS alongside Linux to run NUMA nodes within HermitCore instances which manage all the resources. Further, developers implemented a fast message passing interface realizing an inter-kernel communication between the HermitCore instances. Recently, the authors enables HermitCore to both run (as unikernel) within a VM but also as bare-metal applications \cite{lankes2017low}. In this case, HermitCore could be exploited to run real-time and cloud workloads, since reduced memory footprint and reduced pressure on cache system can provide more predictable behaviors. Further, the authors extended HermitCore to support also many-core architectures. In \cite{lankes2017low}, the authors evaluate HermitCore to reveal the overhead induced on the target system. They leveraged Hourglass benchmark \cite{hourglass, regehr2002inferring} to determine the gaps in the execution time caused by Linux and HermitCore. The results show that HermitCore provides the smallest noise, consequently could be used for real-time scenarios.}

\vspace{2mm}

\begin{mybox}{\textbf{Unikernels}}

$\tikztriangleright[blue,fill=gray!80]$ Fast boot and migration time, low memory footprint, high density, high performance, and an effortless (theoretically) certification process.

$\tikztriangleright[blue,fill=gray!80]$ Leverage the underlying host hypervisor to provide strong security.

$\tikztriangleright[blue,fill=gray!80]$ Applications need to be manually ported to the underlying unikernel.

$\tikztriangleright[blue,fill=gray!80]$ More analysis and studies are needed to assess the feasibility of adoption of these solutions in mixed-criticality real-time systems, especially concerning dependability support and certification.

\end{mybox}


\section{Discussion}
\label{sec:discussion}

\begin{table*}[!ht]
\caption{Comparison between virtualization solutions features through industry dimensions}
\label{tab:summary}
\resizebox{\textwidth}{!}{%
\sffamily 
\footnotesize 
\setstretch{0.90}
\begin{tabular}{>{\centering}m{2.5cm}>{\centering}m{1.8cm}>{\centering}m{2cm}>{\centering}m{1.2cm}>{\centering}m{1.4cm}>{\centering}m{1.1cm}>{\centering}m{2.2cm}>{\centering}m{1.7cm}>{\centering}m{4.0cm}>{\centering}m{4.5cm}}
\cmidrule[4pt]{2-10}

\rowcolor{blue!8}

\cellcolor{white}\multirow{1}{3cm}{} &
\multicolumn{4}{c}{\cellcolor{blue!15}\large\textbf{Solution Features}} &
\multicolumn{3}{c}{\cellcolor{blue!25}\large\textbf{Reuse of legacy}} &  \centering\cellcolor{blue!35}\large\textbf{Dependability} & 
\centering\cellcolor{blue!45}\large\textbf{Certification \& Testing}\tabularnewline
\cmidrule[2pt]{1-10}

\rowcolor{blue!8}
\centering\normalsize\textbf{Category}&
    \cellcolor{blue!15}\centering\normalsize\textbf{Name} & 
    \cellcolor{blue!15}\centering\normalsize\textbf{Hypervisor Type/\revision{Support}} & 
    \cellcolor{blue!15}\centering\normalsize\textbf{\revision{Size}} & 
    \cellcolor{blue!15}\centering\normalsize\textbf{\revision{Latest Release}} & 
    \cellcolor{blue!25}\centering\normalsize\textbf{License} & 
    \cellcolor{blue!25}\centering\normalsize\textbf{Supported Hardware architectures} & 
    \cellcolor{blue!25}\centering\normalsize\textbf{Supported Guest OS/\\\revision{Application}} & 
    \cellcolor{blue!35}\centering\normalsize\textbf{Security, reliability and fault-tolerance features} & 
    \cellcolor{blue!45}\centering\normalsize\textbf{Test suites, test reports and compliance to standards}\tabularnewline
\cmidrule[2pt]{1-10}

& 

    \textbf{VxWorks MILS} & Type-1, Static & Small & \revision{N/A (supersed by VxWorks 653)} & Closed & ARMv7, ARMv8, MIPS, PowerPC, SH, Hitachi H8 & VxWorks Guest OS, WindRiver Linux Guest & Protection from interference and tampering, enforces a Least Privilege Abstraction Partitioned Information Flow Policy (PIFP), supports runtime secure state verification, resource isolation, trusted initialization, trusted delivery, trusted recovery, and audit capabilities & ARINC 653-compliant, MILS support, proven in DO-178C, EUROCAE ED-12, and IEC 61508, Common Criteria with SKPP profile\tabularnewline

 \rowcolor{lightgray!50} \cellcolor{white}
    & \textbf{PikeOS} & Type-1, Dynamic & Small & \revision{v5.0, Feb 2020} & Closed & Intel x86, ARMv8, PowerPC, SPARC V8/LEON, MIPS & Linux, Android, RT-POSIX, ARINC653, RTEMS & Resource isolation, built-in Health Monitoring Function which implements all features described in the ARINC-653 standard, Hardware support for voting (SAFe-VX Architecture) & Compliant to ARINC 653, RTCA DO-178B/C, ISO 26262, IEC 62304, EN 50128, IEC 61508, Common Criteria, SAR, MILS (three security levels). Verification of microkernel system calls \cite{baumann2009verifying}, intransitive noninterfernce properties \cite{verbeek2015formal}\tabularnewline

     & \textbf{NOVA} & Type-1, Dynamic & Medium & \revision{Latest commit Jul 2021} & GPL & Intel x86, x86-64, ARMv8-based boards & Linux & Protection against VMM attacks, guest attacks & Formal verification analysis\cite{tews2008nova}, performance testing and overhead analysis \cite{steinberg2010nova}\tabularnewline

   \rowcolor{lightgray!50} \cellcolor{white}
    & \revision{\textbf{sel4}} & \revision{Type-1, Dynamic} & \revision{Small} & \revision{Latest commit Oct 2021} & \revision{GPL} & \revision{Intel x86, x86-64, ARMv7/v8-based boards, RISC-V} & \revision{Linux} & \revision{Time and space partitioning \cite{lyons2018scheduling}.\\ Proofs of security enforcement \cite{klein2014comprehensive}.\\ Task backup and recovery mechanisms, and checkpoint-based optimization strategies \cite{luan2018towards, xu2016towards}} & \revision{Implementation correctness formal-verified. WCET analysis provided in {\cite{blackham2011timing, sewell2017high}}. Certifiable in theory.} \tabularnewline

    & \textbf{Xtratum} & Type-1, Static & Small & \revision{v1.0.8, Jan 2019} & Both GPL and closed & Intel x86, SPARCv8 (Leon2, Leon3, Leon4), ARMv7 (Cortex R4, R5, A9), PowerPC & Linux, RTEMS, ORK+, LITHOS RTOS, Partikle RTOS+ & Temporal and spatial isolation provided. Xtratum was also used as a basis for a hypervisor-based fault tolerant architecture for space applications, providing error detection mechanism via task-level duplication \cite{campagna_xtratum}. & A solution which includes Xtratum and the RTOS ORK+, has been certified to be compliant to the standard ARINC 653.\tabularnewline

    \rowcolor{lightgray!50} \cellcolor{white} \multirow{-35}{2.5cm}{\centering\textbf{Separation kernel and Microkernel}} & \textbf{Jailhouse} & 
        Type-1, Static & 
        Small & 
        \revision{v0.12, Oct 2021} & 
        GPL & 
        Intel x86, ARMv7 and ARMv8 (32 and 64 bit) & 
        Linux, L4 Fiasco.OC, FreeRTOS, Erika Enterprise RTOS v3 & 
        Resource isolation, deadlock detection and automatic restart & 
        Released applying CI and static code analysis tools. No certification was done; within the SELENE project \cite{selene_project} researchers are working on IEC 61508 compliance.
        \tabularnewline
\toprule\toprule 

 & 

    \textbf{Xen/RT-Xen} & Type-1, Dynamic & Large (x86, x86-64)\\ Medium (ARM) & \revision{Xen v4.15, Apr 2021 - RT-Xen latest commit Jan 2016} & GPL & Intel x86, x86-64, ARMv8 & Linux, Windows & Resource isolation, high availability support with proprietary tools (Citrix) & \textit{Xen Test Framework (XTF)} \cite{xen_test_framework} with assessment of specific security vulnerability, sanity checks, and functional tests. \textit{OSSTest} \cite{osstest_xen} to leverage CI tools. Safety certification efforts provided by \textit{FuSa SIG} \cite{xen_fusa_sig}. \tabularnewline
 
    \rowcolor{lightgray!50} \cellcolor{white} \multirow{-7}{2.5cm}{\centering\textbf{General-purpose}} & \textbf{KVM} & Type-1, Dynamic & Large & \revision{Latest commit Oct 2021, integrated with Linux kernel v5.14} & GPL & Intel x86, x86-64 & Linux, BSD, QNX, Windows & Resource isolation, high availability support with proprietary tools (oVirt) & PREEMPT\_RT is accompanied with real-time test cases within LTP, and benchmarks for worst-case latency scenarios, performance analysis, scheduling attributes tests, priority inversion deadlock test. \tabularnewline
\toprule\toprule 

& 
    \textbf{LTZVisor} & Type-1, Dynamic (DUAL OS RTOS+ GPOS) & Small & \revision{Latest commit Oct 2017} & GPL & ARMv7-A, ARMv8-A and ARMv8-M & Linux, FreeRTOS & Spatial and temporal isolation enfornced by security hardware extensions, for mixed-criticality applications & Test suites for functional testing and performance analysis \cite{pinto2017ltzvisor,pinto2017lightweight}\tabularnewline
 
    \rowcolor{lightgray!50} \cellcolor{white}  & \textbf{RTZVisor/ $\mu$RTZVisor} & Type-1, Dynamic & Small & \revision{N/A} & Closed & ARMv7-A, ARMv8-A and ARMv8-M & Linux, FreeRTOS & Spatial and temporal isolation enfornced by security hardware extensions, for mixed-criticality applications & Test suites for functional testing and performance analysis \cite{pinto2016towards,martins2017murtzvisor}\tabularnewline
 
    \multirow{-10}{2.5cm}{\centering\textbf{ARM TrustZone-assisted}} & \textbf{VOSYSmonitor} & Type-1, Dynamic & Small & \revision{Last website update 2021} & Closed & ARMv7-A, ARMv8-A& Automotive grade Linux, Andorid, AUTOSAR & Spatial and temporal isolation enfornced by security hardware extensions, fault detection and safe software migration. & ISO 26262 compliant at ASIL-C level. Self tests for run-time functional testing and performance analysis.
\tabularnewline
\toprule\toprule 

\rowcolor{lightgray!50} \cellcolor{white} & 

\textbf{Linux Containers / RT-Case} & 
    Process Containers, Dynamic & Medium & \revision{Docker latest commit Oct 2021\\RT-Case latest commit Sept. 2021} & 
    GPL & 
    Intel x86 and x86-64, ARM and many others & 
    Linux applications & Resource isolation, high availability support through Docker tools. Security support inherited from Linux (SELinux and AppArmor frameworks, sandboxing support, Mandatory Access Control). & 
    Docker is provided with Unit test and API Integration test suites. LXC is provided with a continuous security analysis framework. Docker and LXC are released through CI tools.\tabularnewline

& \textbf{ClickOS Unikernel} & 

    Xen-based & 
    Small & \revision{Latest commit Oct 2017. Forked from \textit{click} repo, latest commit Oct 2020} & 
    GPL & 
    Inherit from Xen supported CPUs & 
    Several \textit{Click} \cite{kohler2000click} middleboxes & 
    Strong isolation and security due to unikernel-based design, but no evidence is provided in \cite{martins2014clickos}. & 
    Mention to the certification procesess, but no evidence is provided in \cite{martins2014clickos}.\tabularnewline
    
\rowcolor{lightgray!50} \cellcolor{white} \multirow{-18}{2.5cm}{\centering\textbf{Lightweight Virtualization}} & \revision{\textbf{HermitCore Unikernel}} & 

    \revision{uhyve (KVM-based), KVM, QEMU} & 
    \revision{Medium} & \revision{Latest commit Jul 2020} & 
    \revision{GPL} & 
    \revision{Inherit from hypervisors supported CPUs} & 
    \revision{Applications built via the provided cross toolchain} & 
    \revision{Strong isolation and security due to unikernel-based design, with some empirical results provided in \cite{lankes2016hermitcore, olivier2019binary}.} & 
    \revision{Released applying CI/CD tools.} \tabularnewline
       
    \cmidrule[2pt]{1-10}

\end{tabular}
}
\end{table*}



\tablename~\ref{tab:summary} shows a summary of the main features of each solution. \revision{We considered three classes for hypervisor size according to Lines of Code (LOC), namely, \emph{Small} (less than $10$kLOC), \emph{Medium} (less than $100k$LOC) and \emph{Large} (greater than $100k$LOC) classes. Further, \tablename~\ref{tab:summary} reports for each solution} the license and the supported hardware and guest OS to ease out the porting of existing legacy systems in the virtualization world. Then the table summarizes the dependability features, the availability of test-suites, and the compliance of the product with industry safety and security standards to aid industry practitioners to choose the most appropriate virtualization technology according to their domain needs. In order to aid industry practitioners to choose an appropriate solution to migrate existing legacy systems to a virtualization paradigm, we reported in \tablename~\ref{tab:summary}, for each solution, the kind of license, the supported hardware architectures, and the explicit support to guest OS. 

\revision{Indeed, the \textit{hypervisor selection} is a crucial task in the industrial domain, and it should comply with dimensions we provided in \tablename~\ref{tab:summary} in order to properly migrate to virtualization-based systems. As relevant examples, the HERCULES \cite{hercules2020}, SELENE \cite{selene_project}, and HERMES \cite{hermes_project} H2020 European projects involved several industry partners (e.g., Airbus, Thales Alenia Space, STMicroelectronics, etc.) which cooperate with academia to leverage virtualization technologies in different domain ranging from the railway to aerospace. In that case, the hypervisor selection follows specific requirements that are easily mapped (in some cases directly) to \tablename~\ref{tab:summary}.}



In the following, we provide some points of discussion for state-of-the-practice solution categories reviewed in Section \ref{sec:solutions}, highlighting the current industrial and scientific trends in virtualization.

\vspace{2mm}

\noindent
$\rhd$ \textbf{\textit{Separation kernels and microkernels: the current trend}}. The majority of industry solutions for virtualization fall into the separation kernel and microkernel classes. Also, static approaches to virtualization (aka partitioning) are preferred over dynamic solutions. This is a clever choice since industry scenarios must ensure the highest level of isolation between virtualized domains due to strict requirements that must be met by safety standards clauses. Proprietary solutions (i.e., VxWorks, PikeOS, VOSYSmonitor) support the majority of features required by a safe and secure environment (e.g., run-time secure state verification, health monitoring, trust recovery, etc.). PikeOS also supports the SAFe-VX architecture for voting, which eases the development of reliable applications in safety-critical domains. While open ARM TrustZone-based solutions inherit isolation and security from the underlying hardware, general-purpose and OS-level solutions can take advantage of existing tools developed for supporting high-availability mechanisms in cloud applications (i.e., Citrix HA for Xen, Red Hat oVirt for KVM and Docker tools for the Linux containers).

As one would expect, the most advanced solutions, in terms of certification, are proprietary. VxWorks, PikeOS, VOSYSmonitor are examples of certified solutions, i.e., compliant with the industry safety and security standards, such as ARINC-653, DO-178C, Common criteria, ISO26262, etc... However, recent initiatives in open-source projects are trying to reduce the gap. For instance, Xtratum, with the RTOS ORK+ as a Guest OS, has been certified to be compliant with the ARINC-653 standard.

\vspace{2mm}

\noindent
$\rhd$ \textbf{\textit{General-purpose real-time hypervisors: a new opportunity.}} Also for general-purpose open-source solutions (e.g., KVM, Xen), widely adopted in cloud computing scenarios, we are assisting to a proliferation of projects that are trying to delineate guidelines with tools and methodologies supporting the safety certification process also for these open-source platforms. 

For example, the FuSa Special Interest Group (SiG) is analyzing the possibilities of using Xen as a basis for safety-critical virtualized systems. Indeed, Xen currently provides real-time support for scheduling (ARINC, RTDS, and Null schedulers), a minimal size (less than 30KSLOC) for ARM-based hardware environments, paravirtual and GPU mediation for rich I/O, TEE virtualization support {\cite{xen_fusa_sig}}. \revision{Xen developers provided the \textit{Dom0-less} patch since Xen v4.12 \cite{xen_fusa_sig, dom0less_1, dom0less_2}. This crucial feature enables Xen to create a set of unprivileged domains at boot time, passing information about these VMs to the hypervisor via the \textit{Device Tree} (a tree data structure with nodes that describe the physical devices in Linux-based systems). Indeed, Xen developers extended the older \textit{Device Tree} to allow for multiple domains to be passed to Xen. Actually, the Dom0 is still required to manage the DomUs, but the hypervisor can create additional VMs in parallel without any interactions with the control domain. Practitioners can also omit the definition of Dom0 into the \textit{Device Tree} without specifying the Dom0 kernel obtaining a ``true Dom0-less'' system, but having a Dom0 environment can still be convenient for monitoring and management purposes. “True Dom0-less” configurations fit well scenarios with higher security (reduced attack surface) or to improve resource utilization (shorten boot times). Further, there are several efforts to break into privileged service domains the Dom0 (aka Dom0 disaggregation) to improve security, reliability, and isolation of Xen {\cite{colp2011breaking, dom0_disaggregation, mvondo2020fine}}.} Open-source virtualization solutions are also gaining popularity in the automotive domain, thanks to vertical initiatives such as the Automotive Grade Linux (AGL) project. AGL is considering hypervisors (including Xen, but also OS-level virtualization like Docker) to create a safety-critical execution environment for workloads in software-defined vehicle architecture according to ISO 26262 {\cite{agl_kvm}}. Further, the recent ELISA project {\cite{elisa_project}} promises to implement a certifiable Linux kernel, which (indirectly) impacts KVM applicability for safety use cases.

As mentioned at the beginning, general-purpose solutions fully support cloud computing infrastructures, with several frameworks for the management and orchestration of VMs, which include migration, balancing, and high-availability mechanisms. By leveraging this kind of hypervisors in embedded systems we can easily support the implementation of solutions for orchestrating tasks at different criticality running on different RTOSes and GPOSes within different hardware boards. In the context of the LF Edge foundation {\cite{lf_edge}}, whose goal is to aid the development of industrial IoT and edge devices, Xilinx is currently developing a lightweight solution called RunX, which exploits Xen to both run containers as VMs, \revision{either with the provided custom-built Linux-based kernel with a Busybox-based ramdisk, or with container-specific kernel/ramdisk}.

%


\vspace{2mm}

\noindent
$\rhd$ \textbf{\textit{\revision{TEE-based virtualization}: exploiting hardware-driven innovations.}} ARM TrustZone-based solutions are gaining big momentum today because several embedded system providers build their products on top of ARM CPUs (e.g., Xilinx). However, this kind of virtualization reduces the reuse of legacy software for platforms powered by other CPU vendors like Intel, LEON, and others. In this regard, Intel is supporting the ACRN project, which is an open-source hypervisor with a focus on industrial IoT scenarios and edge device use cases {\cite{li2019acrn}}. 

\vspace{2mm}

\noindent
$\rhd$ \textbf{\textit{Lightweight virtualization: a promising new trend.}} Concerning lightweight virtualization solutions, they are gaining traction for mixed-criticality systems. Especially in the telco industry, we are witnessing the trend to \textit{softwarize} hardware-based network elements towards so-called \textit{virtual network functions} {\cite{nfv_bench}}, for which real-time and mixed-criticality are stringent requirements. Since this kind of virtualization allows delivering low-latency, bandwidth-efficient, and resilient services they fit well use cases like autonomous vehicles, smart cities, and augmented reality, which are common scenarios in industrial IoT {\cite{morabito2018consolidate}}. However, technological questions remain for ensuring reliability and security, but also the timeliness required both for telecommunication networks and mixed-criticality systems.
By using container-based virtualization, the main advantages come with the easy use of built-in orchestration mechanisms (e.g., Docker Swarm) and platforms (e.g., Kubernetes). Containers reduce the overhead affecting VMs and better scale when a larger number of applications of different criticalities are in place with built-in orchestration capabilities. However, containers reduce isolation, threatening the practicability of OS-level virtualization under strict real-time and safety requirements. For example, in {\cite{goldschmidt2018container}} the authors presented an architecture for a multipurpose industrial controller deployed via containers. In {\cite{morabito2017evaluating}}, the authors provide a performance evaluation that aims to show the strengths and weaknesses of different low-power devices when handling container-virtualized instances.

Instead, since unikernel-based solutions do not share the underlying host kernel (each unikernel has its own kernel), they are mainly used to enhance security; furthermore, since unikernels are minimalistic OSes, with image size less than 5 MBS and memory footprint of 8 MBs on average {\cite{morabito2018consolidate}}, this kind of virtualization is a good candidate to ease the certification process for safety-critical mixed-criticality systems. Currently, researchers are exploring unikernel-based solutions in the context of industrial IoT (IIoT) scenarios, which impose critical requirements like determinism, safety certification, isolation, and flexibility. In {\cite{morabito2018consolidate}}, the authors try to understand if unikernels can be exploited for deploying IoT edge architectures and environments, like vehicular cloud computing, edge computing for smart cities, and augmented reality. In {\cite{kurek2019unikernel}}, the authors discuss how to leverage unikernel-based virtualization in the context of NFV IoT gateways. They highlight how the use of containers for NFV could negatively impact security and isolation due to the shared host kernel.

\vspace{2mm}



\noindent
$\rhd$ \textbf{\textit{Hypervisor certification directions.}} Generally, certifying a hypervisor includes several burdensome tasks (e.g., rigorous documentation, test suites, verification tools, and so on) that lead to an increased overall cost of developing safety-critical systems. However, safety standards like EN 50128 and ISO 26262, consider the possibility of integrating \textit{pre-existing software} or into systems being certified. Thus, an interesting research direction is considering a hypervisor as a library to be integrated into a system already certified at some SIL level.

Despite the great maturity of safety-related standards, today there is still a need of facing security certification in the context of new industry movements like IIoT and Industry 4.0. This brings cybersecurity aspects in mixed-criticality systems development, which are not considered in the past. When certifying security with safety there is a need to identify properly the overlap between standards processes and ensure that all security and safety requirements are included, still keeping the overall cost of certification low. These issues are today exacerbated if we consider virtualized mixed-criticality systems.

Further, the use of Machine Learning (ML)/Artificial Intelligence (AI) for bringing autonomy in mixed-criticality applications, where software development shifts from traditional coding to \textit{example-based} training, introduces new issues. \revision{Indeed, several industry domains, especially in automotive and healthcare, currently leverage or plan to use ML/AI techniques for critical decision-making components. Clearly, this leads the developed systems to be ready to co-locate on the same hardware platform non-critical applications (e.g., dashboards, monitoring functions) with highly complex AI components. The SELENE project \cite{selene_project} is a real example of how research and industry are envisioning to apply virtualization technologies (in this specific case, they chose the Jailhouse partitioning hypervisors) to quantify and assess the reliability level that can be reached by placing AI components in a safety-critical system.}

Curating the training process, operating and integrating ML models, and achieving confidence in the ML models through new forms of verification and validation and through "explainable AI" (XAI) techniques, are some of the tasks to be performed. In that case, is crucial to understand how to certify these systems since in safety-related standards it is common to explicitly not recommend using artificial intelligence for almost all safety integrity levels. However, there are improvements in that direction since ISO/IEC provides standards like the ISO/IEC DTR 24029-1 {\cite{iso24029}}, which focuses on the robustness of neural networks, and the ISO/IEC WD TS 4213 {\cite{iso4213}} (under-development), which focuses on the assessment of machine learning classification performance.

\vspace{2mm}

\noindent
$\rhd$ \textbf{\textit{Hybrid virtualization solutions.}} Finally, regardless of the virtualization approaches and solutions analyzed in this paper, we are witnessing trends in adopting hybrid solutions that try to satisfy both real-time and general-purpose needs, simultaneously. This is the case, for instance, of the IoT domain in which there is a need for high portability and adaptability, with a rich set of I/O virtualization capabilities. Virtualization will be extensively used also in high-performance computing (HPC) platforms, which offer the power needed by the modern industrial systems and edge computing architectures by using devices like GPUs, FPGAs, and other kinds of accelerators. No solutions are available yet, capable to face such heterogeneous environments while guaranteeing easy porting, isolation, and real-time properties. Containers are a promising solution for these contexts, combining flexibility and scalability, but they are not mature yet for full adoption in industrial domains.


\section{Conclusion}
\label{sec:conclusion}

This survey analyzed the most important virtualization approaches and related solutions proposed in the last years targeting the real-time and/or safety-critical domains. In particular, we analyzed existing solutions along three fundamental dimensions, which reflect the most common requirements in mixed-criticality domains: \textit{Certification \& Testing}, \textit{Reuse of legacy systems}, and \textit{Dependability support}.

We observed that \textit{separation kernel solutions} are designed to comply with safety certification, and with a high level of isolation, whereas \textit{microkernels approach} provides strong security and effective verification by reducing at the minimum the Trusted Computing Base (TCB). 
Although the previous considerations, \textit{general-purpose solutions} are still a good choice for real-time purposes, and recent initiatives are emerging, to foster their adoption in safety-critical scenarios. 
\textit{Hardware-assisted solutions}, based on \emph{ARM TrustZone}, leverages security features from the hardware and provide well-defined test suites, performance analysis tools, and failure recovery mechanisms. These solutions raise the portability problem on other platforms, and migrating legacy applications may require non-negligible costs.
Finally, \emph{lightweight solutions} are a recent trend, particularly promising to overcome footprint issues while assuring the isolation required for mixed-criticality. However, their adoption in industrial domains, apart from telecommunication networks, is still far to be established. 

More research efforts are needed in several directions. Regarding testing and certification, many safety and security standards provide guidelines for testing activities, which encompass fault injection testing, robustness testing, and performance testing, along with the classical testing activities. However, we still witness a lack of shared benchmarks and effective test suites that could help to produce evidence to support the certification process, especially concerning novel trends, such as the use of lightweight virtualization or the certification of systems based on machine learning and artificial intelligence.
Finally, given the evolution of existing security standards, like ISO 62443 and its derivative EN 50701 for the railway, a good portion of mixed-criticality systems will also require security and privacy certification, which is neglected by most of the current non-commercial solutions.

\Urlmuskip=0mu plus 1mu\relax

\bibliography{bibliography,bibliography_2}

\begin{thebibliography}{100}
\expandafter\ifx\csname url\endcsname\relax
  \def\url#1{\texttt{#1}}\fi
\expandafter\ifx\csname urlprefix\endcsname\relax\def\urlprefix{URL }\fi
\expandafter\ifx\csname href\endcsname\relax
  \def\href#1#2{#2} \def\path#1{#1}\fi

\bibitem{shift2rail}
{Shift2Rail}, {Home page of Shift2Rail projects},
  \url{https://projects.shift2rail.org/s2r_projects.aspx}.

\bibitem{windriveriot}
{WindRiver Systems Inc.},
  \href{https://www.windriver.com/whitepapers/iot-virtualization/1436-IoT-Virtualization-White-Paper.pdf}{{Virtualization
  and the Internet of Things}}, WindRiver White Paper (2016) 4.
\newline\urlprefix\url{https://www.windriver.com/whitepapers/iot-virtualization/1436-IoT-Virtualization-White-Paper.pdf}

\bibitem{hercules2020}
{Hercules 2020}, {Home page of Hercules 2020 project},
  \url{http://hercules2020.eu/}.

\bibitem{5gcity_H2020}
{5GCityl}, {5GCity Project Home Page}, \url{https://www.5gcity.eu/}.

\bibitem{hermann2016design}
M.~Hermann, T.~Pentek, B.~Otto, Design principles for industrie 4.0 scenarios,
  in: Proc. HICSS, IEEE, 2016, pp. 3928--3937.

\bibitem{klingensmith2019using}
N.~Klingensmith, S.~Banerjee, Using virtualized task isolation to improve
  responsiveness in mobile and iot software, in: Proc. IoTDI, ACM/IEEE, 2019,
  pp. 160--171.

\bibitem{klingensmith2018hermes}
N.~Klingensmith, S.~Banerjee, Hermes: A real time hypervisor for mobile and iot
  systems, in: Proc. HotMobile, ACM, 2018, pp. 101--106.

\bibitem{heiser2011virtualizing}
G.~Heiser, Virtualizing embedded systems-why bother?, in: 2011 48th
  ACM/EDAC/IEEE Design Automation Conference (DAC), IEEE, 2011, pp. 901--905.

\bibitem{do178b}
RTCA, {DO-178B Software Considerations in Airborne Systems and Equipment
  Certification}, Requirements and Technical Concepts for Aviation.

\bibitem{iso26262}
{ISO}, {Product Development: Software Level}, ISO 26262: Road vehicles --
  Functional safety 6.

\bibitem{cenelec201150128}
{CENELEC}, {EN 50128}, Railway applications-Communication, Signaling and
  Processing Systems-Software for Railway Control and Protection Systems.

\bibitem{garcia2014challenges}
M.~Garc{\'\i}a-Valls, T.~Cucinotta, C.~Lu, {Challenges in Real-Time
  Virtualization and Predictable Cloud Computing}, Elsevier JSA 60~(9) (2014)
  726--740.

\bibitem{gu2012state}
Z.~Gu, Q.~Zhao, A state-of-the-art survey on real-time issues in embedded
  systems virtualization, Scientific Research Publishing Journal of Software
  Engineering and Applications 5~(4) (2012) 277--290.

\bibitem{burns2018survey}
A.~Burns, R.~I. Davis, {A Survey of Research into Mixed Criticality Systems},
  ACM CSUR 50~(6) (2018) 82.

\bibitem{taccari2014embedded}
G.~Taccari, L.~Taccari, A.~Fioravanti, L.~Spalazzi, A.~Claudi, A.~B. SA,
  Embedded real-time virtualization: State of the art and research challenges,
  in: Proc. RTLWS, 2014, pp. 1--7.

\bibitem{reghenzani2019real}
F.~Reghenzani, G.~Massari, W.~Fornaciari, The real-time linux kernel: A survey
  on preempt\_rt, ACM CSUR 52~(1) (2019) 1--36.

\bibitem{Struhar5796}
V.~Struhar, M.~Behnam, M.~Ashjaei, A.~Papadopoulos, Real-time containers: A
  survey, in: Proc. Fog-IoT Workshop, 2020.

\bibitem{vmware_esxi}
{VMware Inc.},
  \href{http://www.vmware.com/it/products/esxi-and-esx/overview.html}{{VMware
  ESXi Overview}}.
\newline\urlprefix\url{http://www.vmware.com/it/products/esxi-and-esx/overview.html}

\bibitem{kivity07kvm}
A.~Kivity, Y.~Kamay, D.~Laor, U.~Lublin, A.~Liguori, kvm: the linux virtual
  machine monitor, in: Proc. Linux Symp., Vol.~1, 2007, pp. 225--230.

\bibitem{microsoft_hyperv}
{Microsoft Corporation},
  \href{http://technet.microsoft.com/en-us/windowsserver/dd448604.aspx}{{Hyper-V}}.
\newline\urlprefix\url{http://technet.microsoft.com/en-us/windowsserver/dd448604.aspx}

\bibitem{XEN_Barham2003}
P.~Barham, B.~Dragovic, K.~Fraser, S.~Hand, T.~Harris, A.~Ho, R.~Neugebauer,
  I.~Pratt, A.~Warfield, {Xen and the art of virtualization}, in: Proc. SOSP,
  2003, pp. 164--177.

\bibitem{heiser2008role}
G.~Heiser, The role of virtualization in embedded systems, in: Proc. IIES,
  2008, pp. 11--16.

\bibitem{runx}
{Xilinx}, {RunX}, \url{https://github.com/Xilinx/runx}.

\bibitem{kubernetes_frakti}
Frakti, {Frakti GitHub page}, \url{https://github.com/kubernetes/frakti}.

\bibitem{bugnion2012bringing}
E.~Bugnion, S.~Devine, M.~Rosenblum, J.~Sugerman, E.~Y. Wang, {Bringing
  Virtualization to the x86 Architecture with the Original VMware Workstation},
  ACM TOCS 30~(4).

\bibitem{inject_hw_fault_hypervisor}
M.~Cinque, A.~Pecchia, {On the Injection of Hardware Faults in Virtualized
  Multicore Systems}, {Elsevier Journal of Parallel and Distributed Computing}
  106 (2017) 50--61.

\bibitem{jailhouse_perf_isolation_test}
J.~{Danielsson}, T.~{Seceleanu}, M.~{Jägemar}, M.~{Behnam}, M.~{Sjödin},
  {Testing Performance-Isolation in Multi-core Systems}, in: Proc. COMPSAC,
  2019, pp. 604--609.

\bibitem{iec61508}
{International Electrotechnical Commission}, {Software Requirements}, IEC
  61508-3.

\bibitem{arinc653}
{Aeronautical Radio Inc.}, {ARINC-653: Avionics application Software standard
  interface part 1} (2010).

\bibitem{CAST-32A}
{Certification Authorities Software Team (CAST)}, {Multi-core Processors},
  \url{https://www.faa.gov/aircraft/air_cert/design_approvals/air_software/cast/cast_papers/media/cast-32a.pdf}.

\bibitem{cotroneo2013fault}
D.~Cotroneo, R.~Natella, {Fault Injection for Software Certification}, IEEE
  Security \& Privacy 11~(4) (2013) 38--45.

\bibitem{cotroneo2012experimental}
D.~Cotroneo, A.~Lanzaro, R.~Natella, R.~Barbosa, {Experimental Analysis of
  Binary-level Software Fault Injection in Complex Software}, in: Proc. EDCC,
  IEEE, 2012, pp. 162--172.

\bibitem{cotroneo2018run}
D.~Cotroneo, L.~De~Simone, R.~Natella, {Run-Time Detection of Protocol Bugs in
  Storage I/O Device Drivers}, IEEE TR 67~(3) (2018) 847--869.

\bibitem{cotroneo2016faultprog}
D.~Cotroneo, A.~Lanzaro, R.~Natella, {Faultprog: Testing the Accuracy of
  Binary-level Software Fault Injection}, IEEE TDSC 15~(1) (2016) 40--53.

\bibitem{nfv_dep_guidelines}
D.~{Cotroneo}, L.~{De Simone}, R.~{Natella}, {Dependability Certification
  Guidelines for NFVIs through Fault Injection}, in: Proc. ISSREW, IEEE, 2018,
  pp. 321--328.

\bibitem{winter2015no}
S.~Winter, O.~Schwahn, R.~Natella, N.~Suri, D.~Cotroneo, {No PAIN, no gain?:
  the utility of PArallel fault INjections}, in: Proc. ICSE, IEEE Press, 2015,
  pp. 494--505.

\bibitem{mazzeo2018sil2}
G.~Mazzeo, L.~Coppolino, S.~D’Antonio, C.~Mazzariello, L.~Romano, Sil2
  assessment of an active/standby cots-based safety-related system, Elsevier
  Reliability Engineering \& System Safety 176 (2018) 125--134.

\bibitem{nasastd2004}
NASA, {Software Safety Guidebook}, NASA-GB-8719.13.

\bibitem{ISO/IEC25045}
{RTcA, RTCA DO}, {ISO/IEC 25045}, Systems and Software Engineering - Systems
  and Software Quality Requirements and Evaluation (SQuaRE) - Evaluation module
  for recoverability.

\bibitem{commoncriteria}
ISO/IEC, {Common Criteria for Information Technology Security Evaluation
  (Version 3.1, Revision 4) Part 1-3 (ISO/IEC 15408)} (2012).

\bibitem{skpp}
{Information Assurance Directorate}, {US Government Protection Profile for
  Separation Kernels in Environments Requiring High Robustness}, Tech. rep.,
  National Security Agency (2007).

\bibitem{zhao2017survey}
Y.~Zhao, Z.~Yang, D.~Ma, A survey on formal specification and verification of
  separation kernels, Springer Frontiers of Computer Science 11~(4) (2017)
  585--607.

\bibitem{NIST.SP.800-53r4}
{NIST}, {Security and Privacy Controls for Federal Information Systems and
  Organizations (NIST SP 800-53 R4)},
  \url{http://dx.doi.org/10.6028/NIST.SP.800-53r4.} (2013).

\bibitem{prio_inversion}
{The Linux Foundation}, {Priority inversion - priority inheritance},
  \url{https://wiki.linuxfoundation.org/realtime/documentation/technical_basics/pi}.

\bibitem{lockholder_problem}
B.~Teabe, V.~Nitu, A.~Tchana, D.~Hagimont, {The Lock Holder and the Lock Waiter
  Pre-Emption Problems: Nip Them in the Bud Using Informed Spinlocks
  (I-Spinlock)}, in: Proc. EuroSys, ACM, 2017, p. 286–297.

\bibitem{alves2006mils}
J.~Alves-Foss, P.~W. Oman, C.~Taylor, W.~S. Harrison, {The MILS architecture
  for high-assurance embedded systems}, Inderscience International Journal of
  Embedded Systems 2~(3-4) (2006) 239--247.

\bibitem{vxworks_mils}
{Wind River Systems, Inc.}, {Wind River VxWorks MILS Platform 3.0, multi-core
  edition},
  \url{https://www.windriver.com/products/product-notes/vxworks-mils-multi-core-platform-product-note/vxworks-mils-multi-core-platform-product-note.pdf}.

\bibitem{cotroneo2021timing}
D.~Cotroneo, L.~De~Simone, R.~Natella, Timing covert channel analysis of the
  vxworks mils embedded hypervisor under the common criteria security
  certification, Computers \& Security 106 (2021) 102307.

\bibitem{aroca2009real}
R.~V. Aroca, G.~Caurin, S.~Carlos-SP-Brasil, A real time operating systems
  (rtos) comparison, in: WSO-Workshop de Sistemas Operacionais, Vol.~12,
  Citeseer, 2009.

\bibitem{pikeos}
{PikeOS},
  \href{https://www.sysgo.com/fileadmin/user_upload/www.sysgo.com/redaktion/downloads/pdf/data-sheets/SYSGO-Product-Overview-PikeOS.pdf}{{PikeOS
  product overview}}.
\newline\urlprefix\url{https://www.sysgo.com/fileadmin/user_upload/www.sysgo.com/redaktion/downloads/pdf/data-sheets/SYSGO-Product-Overview-PikeOS.pdf}

\bibitem{august2014idp}
M.~August, {IDP: An Analysis of a Cache-Based Timing Side Channel Attack and a
  Countermeasure on PikeOS} (2014).

\bibitem{heron2009advanced}
S.~Heron, {Advanced encryption standard (AES)}, Elsevier Network Security
  2009~(12) (2009) 8--12.

\bibitem{verbeek2015formal}
F.~Verbeek, O.~Havle, J.~Schmaltz, S.~Tverdyshev, H.~Blasum, B.~Langenstein,
  W.~Stephan, B.~Wolff, Y.~Nemouchi, Formal api specification of the pikeos
  separation kernel, in: NASA Formal Methods Symposium, Springer, 2015, pp.
  375--389.

\bibitem{roscoe1999intransitive}
A.~W. Roscoe, M.~H. Goldsmith, What is intransitive noninterference?, in: Proc.
  CSF Workshop, IEEE, 1999, pp. 228--238.

\bibitem{baumann2009verifying}
C.~Baumann, T.~Bormer, Verifying the pikeos microkernel: first results in the
  verisoft xt avionics project, in: Proc. SSV, 2009, p.~20.

\bibitem{muttillo2019benchmarking}
V.~Muttillo, L.~Tiberi, L.~Pomante, P.~Serri, Benchmarking analysis and
  characterization of hypervisors for space multicore systems, Journal of
  Aerospace Information Systems 16~(11) (2019) 500--511.

\bibitem{masmano2009xtratum}
M.~Masmano, I.~Ripoll, A.~Crespo, J.~Metge, Xtratum: a hypervisor for safety
  critical embedded systems, in: Proc. RTLWS, Citeseer, 2009, pp. 263--272.

\bibitem{zamorano2010open}
J.~Zamorano, J.~de~la Puente, Open source implementation of hierarchical
  scheduling for integrated modular avionics, in: Proc. RTLWS, 2010.

\bibitem{esquinas2011ork}
{\'A}.~Esquinas, J.~Zamorano, A.~Juan, M.~Masmano, I.~Ripoll, A.~Crespo,
  {ORK+/XtratuM: An open partitioning platform for Ada}, in: Proc. Ada-Europe,
  Springer, 2011, pp. 160--173.

\bibitem{oversee}
N.~McGuire, A.~Platschek, G.~Schiesser, Oversee—a generic floss communication
  and application platform for vehicles, in: Proc. RTLWS, 2010.

\bibitem{campagna_xtratum}
S.~Campagna, M.~Hussain, M.~Violante, Hypervisor-based virtual hardware for
  fault tolerance in cots processors targeting space applications, in: Proc.
  DFT, IEEE, 2010, pp. 44--51.

\bibitem{zhouinitial}
R.~Zhou, S.~Bai, B.~Wang, N.~McGuire, Q.~Zhou, L.~Li, Initial performance study
  of xtratum/ppc.

\bibitem{carrascosa2014xtratum}
E.~Carrascosa, J.~Coronel, M.~Masmano, P.~Balbastre, A.~Crespo, Xtratum
  hypervisor redesign for leon4 multicore processor, ACM SIGBED Review 11~(2)
  (2014) 27--31.

\bibitem{jailhouse}
{Siemens AG}, \href{https://github.com/siemens/jailhouse}{{Jailhouse hypervisor
  source code}}.
\newline\urlprefix\url{https://github.com/siemens/jailhouse}

\bibitem{ivshmem}
{QEMU}, {IVSHMEM Documentation page},
  \url{https://www.qemu.org/docs/master/system/ivshmem.html}.

\bibitem{qemu}
{QEMU}, \href{https://www.qemu.org/}{{Homepage of QEMU}}.
\newline\urlprefix\url{https://www.qemu.org/}

\bibitem{selene_project}
{CORDIS}, {SELENE: Self-monitored Dependable platform for High-Performance
  Safety-Critical Systems}, \url{https://cordis.europa.eu/project/id/871467}.

\bibitem{steinberg2010nova}
U.~Steinberg, B.~Kauer, {NOVA: a microhypervisor-based secure virtualization
  architecture}, in: Proc. EuroSys, ACM, 2010, pp. 209--222.

\bibitem{tews2008nova}
H.~Tews, T.~Weber, M.~V{\"o}lp, E.~Poll, M.~van Eekelen, P.~van Rossum, Nova
  micro--hypervisor verification (2008).

\bibitem{klein2009sel4}
G.~Klein, K.~Elphinstone, G.~Heiser, J.~Andronick, D.~Cock, P.~Derrin,
  D.~Elkaduwe, K.~Engelhardt, R.~Kolanski, M.~Norrish, et~al., sel4: Formal
  verification of an os kernel, in: Proceedings of the ACM SIGOPS 22nd
  symposium on Operating systems principles, 2009, pp. 207--220.

\bibitem{elphinstone2013l3}
K.~Elphinstone, G.~Heiser, From l3 to sel4 what have we learnt in 20 years of
  l4 microkernels?, in: Proceedings of the Twenty-Fourth ACM Symposium on
  Operating Systems Principles, 2013, pp. 133--150.

\bibitem{heiser2019can}
G.~Heiser, G.~Klein, T.~Murray, Can we prove time protection?, in: Proceedings
  of the Workshop on Hot Topics in Operating Systems, 2019, pp. 23--29.

\bibitem{lyons2018scheduling}
A.~Lyons, K.~McLeod, H.~Almatary, G.~Heiser, Scheduling-context capabilities: A
  principled, light-weight operating-system mechanism for managing time, in:
  Proc. EuroSys, ACM, 2018, pp. 1--16.

\bibitem{camkes}
{The Linux Foundation},
  \href{https://docs.sel4.systems/projects/camkes/}{{CAmkES documentation}}.
\newline\urlprefix\url{https://docs.sel4.systems/projects/camkes/}

\bibitem{blackham2011timing}
B.~Blackham, Y.~Shi, S.~Chattopadhyay, A.~Roychoudhury, G.~Heiser, Timing
  analysis of a protected operating system kernel, in: 2011 IEEE 32nd Real-Time
  Systems Symposium, IEEE, 2011, pp. 339--348.

\bibitem{sewell2017high}
T.~Sewell, F.~Kam, G.~Heiser, High-assurance timing analysis for a
  high-assurance real-time operating system, Real-Time Systems 53~(5) (2017)
  812--853.

\bibitem{luan2018towards}
G.~Luan, Y.~Bai, L.~Xu, C.~Yu, C.~Wang, J.~Zeng, Q.~Chen, W.~Wang, Towards
  fault-tolerant task backup and recovery in the sel4 microkernel, in: Proc.
  COMPSAC, Vol.~1, IEEE, 2018, pp. 721--726.

\bibitem{xu2016towards}
L.~Xu, Y.~Bai, K.~Cheng, L.~Ge, D.~Nie, L.~Zhang, W.~Liu, Towards
  fault-tolerant real-time scheduling in the sel4 microkernel, in: Proc.
  HPCC/SmartCity/DSS, IEEE, 2016, pp. 711--718.

\bibitem{neiger2006vtx}
G.~Neiger, A.~Santoni, F.~Leung, D.~Rodgers, R.~Uhlig, Intel virtualization
  technology: Hardware support for efficient processor virtualization., Intel
  Technology Journal 10~(3).

\bibitem{abeni2020using}
L.~Abeni, D.~Faggioli, {Using Xen and KVM as Real-Time Hypervisors}, Elsevier
  JSA (2020) 101709.

\bibitem{abeni2019experimental}
L.~Abeni, D.~Faggioli, An experimental analysis of the xen and kvm latencies,
  in: Proc. ISORC, IEEE, 2019, pp. 18--26.

\bibitem{real_time_kvm}
P.~Bonzini, {Realtime KVM}, \url{https://lwn.net/Articles/656807/}.

\bibitem{xi2015rt}
S.~Xi, C.~Li, C.~Lu, C.~D. Gill, M.~Xu, L.~T. Phan, I.~Lee, O.~Sokolsky,
  Rt-open stack: Cpu resource management for real-time cloud computing, in:
  Proc. CLOUD, IEEE, 2015, pp. 179--186.

\bibitem{agl_kvm}
{The Linux Foundation},
  \href{https://www.automotivelinux.org/wp-content/uploads/sites/4/2018/06/agl_software_defined_car_jun18.pdf}{{The
  Automotive Grade Linux Software Defined Connected Car Architecture}}, White
  Paper.
\newline\urlprefix\url{https://www.automotivelinux.org/wp-content/uploads/sites/4/2018/06/agl_software_defined_car_jun18.pdf}

\bibitem{RT-XEN_xi2011rt}
S.~Xi, J.~Wilson, C.~Lu, C.~Gill, Rt-xen: Towards real-time hypervisor
  scheduling in xen, in: Proc. EMSOFT, ACM, 2011, pp. 39--48.

\bibitem{RT-XEN_xi2014real}
S.~Xi, M.~Xu, C.~Lu, L.~T. Phan, C.~Gill, O.~Sokolsky, I.~Lee, Real-time
  multi-core virtual machine scheduling in xen, in: Proc. EMSOFT, IEEE, 2014,
  pp. 1--10.

\bibitem{jeong2011parfait}
J.-W. Jeong, S.~Yoo, C.~Yoo, Parfait: A new scheduler framework supporting
  heterogeneous xen-arm schedulers, in: Proc. CCNC, IEEE, 2011, pp. 1192--1196.

\bibitem{gupta2006enforcing}
D.~Gupta, L.~Cherkasova, R.~Gardner, A.~Vahdat, Enforcing performance isolation
  across virtual machines in xen, in: Proc. Middleware, Springer, 2006, pp.
  342--362.

\bibitem{govindan2009xen}
S.~Govindan, J.~Choi, A.~R. Nath, A.~Das, B.~Urgaonkar, A.~Sivasubramaniam, Xen
  and co.: Communication-aware cpu management in consolidated xen-based hosting
  platforms, IEEE TOC~(8) (2009) 1111--1125.

\bibitem{xen_ultrascale}
{Xilinx, Inc.}, {Enabling Virtualization with Xen Hypervisor on Zynq
  UltraScale+ MPSoCs (White Paper)}, Xilinx.

\bibitem{xen_test_framework}
{Citrix Systems}, {Xen Test Framework Home Page},
  \url{http://xenbits.xen.org/gitweb/?p=osstest.git;a=blob;f=README}.

\bibitem{osstest_xen}
{Citrix Systems}, {OSStest Xen Project REAME},
  \url{http://xenbits.xen.org/gitweb/?p=osstest.git;a=blob;f=README}.

\bibitem{vanderleest2013safe}
S.~H. VanderLeest, D.~Greve, P.~Skentzos, A safe \& secure arinc 653
  hypervisor, in: Proc. DASC, IEEE, 2013, pp. 7B4--1.

\bibitem{xen_fusa_sig}
{FuSa SIG}, {FuSa SIG Charted},
  \url{https://wiki.xen.org/wiki/FwwuSa_SIG/Charter}.

\bibitem{preempt_rt}
P.~McKenney, {A realtime preemption overview},
  \url{https://lwn.net/Articles/146861/}.

\bibitem{kiszka2009towards}
J.~Kiszka, Towards linux as a real-time hypervisor, in: Proc. RTLWS, Citeseer,
  2009, pp. 215--224.

\bibitem{cucinotta2009respecting}
T.~Cucinotta, G.~Anastasi, L.~Abeni, Respecting temporal constraints in
  virtualised services, in: Proc. COMPSAC, Vol.~2, IEEE, 2009, pp. 73--78.

\bibitem{linux_cgroups}
P.~Menage, {cgroups documentation},
  \url{https://www.kernel.org/doc/Documentation/cgroup-v2.txt}.

\bibitem{cucinotta2010providing}
T.~Cucinotta, D.~Giani, D.~Faggioli, F.~Checconi, Providing performance
  guarantees to virtual machines using real-time scheduling, in: Proc.
  Euro-Par, Springer, 2010, pp. 657--664.

\bibitem{zhang2010performance}
J.~Zhang, K.~Chen, B.~Zuo, R.~Ma, Y.~Dong, H.~Guan, {Performance analysis
  towards a KVM-based embedded real-time virtualization architecture}, in:
  Proc. ICCIT, IEEE, 2010, pp. 421--426.

\bibitem{ltp_rt_tests}
{LTP developers}, {Description of LTP real-time test cases},
  \url{https://github.com/linux-test-project/ltp/blob/master/testcases/realtime/00_Descriptions.txt}.

\bibitem{trustzone}
ARM, {TrustZone Technology for Microcontrollers},
  \url{https://www.arm.com/why-arm/technologies/trustzone-for-cortex-m}.

\bibitem{intel_sgx}
V.~Costan, S.~Devadas, Intel sgx explained, Cryptology ePrint Archive, Report
  2016/086, \url{http://eprint.iacr.org/2016/086} (2016).

\bibitem{de2019isolating}
L.~De~Simone, G.~Mazzeo, Isolating real-time safety-critical embedded systems
  via sgx-based lightweight virtualization, in: Proc. ISSREW, IEEE, 2019, pp.
  308--313.

\bibitem{madhavapeddy2013unikernels}
A.~Madhavapeddy, D.~J. Scott, {Unikernels: Rise of the Virtual Library
  Operating System}, ACM Queue 11~(11) (2013) 30.

\bibitem{madhavapeddy2013unikernels_2}
A.~Madhavapeddy, R.~Mortier, C.~Rotsos, D.~Scott, B.~Singh, T.~Gazagnaire,
  S.~Smith, S.~Hand, J.~Crowcroft, Unikernels: Library operating systems for
  the cloud, ACM SIGARCH Computer Architecture News 41~(1) (2013) 461--472.

\bibitem{douglas2010thin}
H.~Douglas, {Thin Hypervisor-based Security Architectures for Embedded
  Platforms}, Ph.D. thesis, Royal Institute of Technology (2010).

\bibitem{frenzel2010arm}
T.~Frenzel, A.~Lackorzynski, A.~Warg, H.~H{\"a}rtig, {ARM TrustZone as a
  Virtualization Technique in Embedded Systems}, in: Proc. RTLWS, 2010, pp.
  29--42.

\bibitem{pinto2017ltzvisor}
S.~Pinto, J.~Pereira, T.~Gomes, A.~Tavares, J.~Cabral, Ltzvisor: Trustzone is
  the key, in: Proc. ECRTS, Schloss Dagstuhl-Leibniz-Zentrum fuer Informatik,
  2017.

\bibitem{oh2012acceleration}
S.-C. Oh, K.~Koh, C.-Y. Kim, K.~Kim, S.~Kim, {Acceleration of dual OS
  virtualization in embedded systems}, in: Proc. ICCCT, IEEE, 2012, pp.
  1098--1101.

\bibitem{schwarz2014affordable}
O.~Schwarz, C.~Gehrmann, V.~Do, Affordable separation on embedded platforms,
  in: Proc. TRUST, Springer, 2014, pp. 37--54.

\bibitem{pinto2017lightweight}
S.~Pinto, A.~Oliveira, J.~Pereira, J.~Cabral, J.~Monteiro, A.~Tavares,
  Lightweight multicore virtualization architecture exploiting arm trustzone,
  in: Proc. IECON, IEEE, 2017, pp. 3562--3567.

\bibitem{pinto2016towards}
S.~Pinto, J.~Pereira, T.~Gomes, M.~Ekpanyapong, A.~Tavares, Towards a
  trustzone-assisted hypervisor for real-time embedded systems, IEEE Computer
  Architecture Letters 16~(2) (2016) 158--161.

\bibitem{martins2017murtzvisor}
J.~Martins, J.~Alves, J.~Cabral, A.~Tavares, S.~Pinto, {$\mu$RTZVisor: A secure
  and safe real-time hypervisor}, MDPI Electronics 6~(4) (2017) 93.

\bibitem{Lucas2018vosys}
P.~Lucas, K.~Chappuis, B.~Boutin, J.~Vetter, D.~Raho, {VOSYSmonitor, a
  TrustZone-based Hypervisor for ISO 26262 Mixed-critical System}, in: Proc.
  FRUCT, IEEE, 2018, pp. 231--238.

\bibitem{lucas2017vosysmonitor}
P.~Lucas, K.~Chappuis, M.~Paolino, N.~Dagieu, D.~Raho, {VOSYSmonitor, a low
  latency monitor layer for mixed-criticality systems on ARMv8-A}, in: Proc
  ECRTS, Schloss Dagstuhl-Leibniz-Zentrum fuer Informatik, 2017.

\bibitem{arm_trusted_firmware}
{ARM Holdings}, {ARM Trusted Firmware repository},
  \url{https://github.com/ARM-software/arm-trusted-firmware}.

\bibitem{rt-containers2020}
V.~Struh{\'a}r, M.~Behnam, M.~Ashjaei, A.~V. Papadopoulos, {Real-Time
  Containers: A Survey}, in: 2nd Workshop on Fog Computing and the IoT (Fog-IoT
  2020), OpenAccess Series in Informatics (OASIcs), 2020.

\bibitem{man2018namespaces}
{Manual, Linux Programmer’s}, \href{http://man7.
  org/linux/man-pages/man7/namespaces}{{Namespaces (7)}}.
\newline\urlprefix\url{http://man7. org/linux/man-pages/man7/namespaces}

\bibitem{cinque2019rt}
M.~Cinque, R.~Della~Corte, A.~Eliso, A.~Pecchia, Rt-cases: Container-based
  virtualization for temporally separated mixed-criticality task sets, in:
  Proc. ECRTS, Schloss Dagstuhl-Leibniz-Zentrum fuer Informatik, 2019.

\bibitem{abeni2019container}
L.~Abeni, A.~Balsini, T.~Cucinotta, Container-based real-time scheduling in the
  linux kernel, ACM SIGBED Review 16~(3) (2019) 33--38.

\bibitem{cucinotta2018virtual}
T.~Cucinotta, L.~Abeni, M.~Marinoni, A.~Balsini, C.~Vitucci, Virtual network
  functions as real-time containers in private clouds., in: IEEE CLOUD, 2018,
  pp. 916--919.

\bibitem{cucinotta2019reducing}
T.~Cucinotta, L.~Abeni, M.~Marinoni, A.~Balsini, C.~Vitucci, Reducing temporal
  interference in private clouds through real-time containers, in: Proc. EDGE,
  IEEE, 2019, pp. 124--131.

\bibitem{nfv_bench}
D.~{Cotroneo}, L.~{De Simone}, R.~{Natella}, {NFV-Bench: A Dependability
  Benchmark for Network Function Virtualization Systems}, IEEE TNSM 14~(4)
  (2017) 934--948.

\bibitem{lxc}
LXC, {LXC - Linux Containers}, \url{https://linuxcontainers.org/}.

\bibitem{olivier2019binary}
P.~Olivier, D.~Chiba, S.~Lankes, C.~Min, B.~Ravindran, A binary-compatible
  unikernel, in: Proceedings of the 15th ACM SIGPLAN/SIGOPS International
  Conference on Virtual Execution Environments, 2019, pp. 59--73.

\bibitem{olivier2021syscall}
P.~Olivier, H.~Lefeuvre, D.~Chiba, S.~Lankes, C.~Min, B.~Ravindran, A
  syscall-level binary-compatible unikernel, IEEE Transactions on Computers.

\bibitem{kuenzer2021unikraft}
S.~Kuenzer, V.-A. B{\u{a}}doiu, H.~Lefeuvre, S.~Santhanam, A.~Jung, G.~Gain,
  C.~Soldani, C.~Lupu, {\c{S}}.~Teodorescu, C.~R{\u{a}}ducanu, et~al.,
  Unikraft: fast, specialized unikernels the easy way, in: Proc. EuroSys, ACM,
  2021, pp. 376--394.

\bibitem{mirageos}
{The Linux Foundation}, {MirageOS Homepage},
  \href{https://mirage.io/}{\nolinkurl{https://mirage.io/}}.

\bibitem{kivity2014osv}
A.~Kivity, D.~Laor, G.~Costa, P.~Enberg, N.~Har’El, D.~Marti, V.~Zolotarov,
  {OSv—optimizing the operating system for virtual machines}, in: Proc.
  USENIX ATC, 2014, pp. 61--72.

\bibitem{rumprun}
{Rumprun developers}, {Rumprun GitHub Homepage},
  \href{https://github.com/rumpkernel/rumprun}{\nolinkurl{https://github.com/rumpkernel/rumprun}}.

\bibitem{wick2012halvm}
A.~Wick, The halvm: A simple platform for simple platforms, Xen Summit.

\bibitem{martins2014clickos}
J.~Martins, M.~Ahmed, C.~Raiciu, V.~Olteanu, M.~Honda, R.~Bifulco, F.~Huici,
  Clickos and the art of network function virtualization, in: Proc. NSDI, 2014,
  pp. 459--473.

\bibitem{popuri2014tour}
S.~Popuri, A tour of the mini-os kernel,
  \url{https://www.cs.uic.edu/~spopuri/minios.html}.

\bibitem{xen_networking}
{The Linux Foundation}, {Xen Networking},
  \url{https://wiki.xenproject.org/wiki/Xen_Networking}.

\bibitem{lankes2016hermitcore}
S.~Lankes, S.~Pickartz, J.~Breitbart, Hermitcore: a unikernel for extreme scale
  computing, in: Proceedings of the 6th International Workshop on Runtime and
  Operating Systems for Supercomputers, 2016, pp. 1--8.

\bibitem{lankes2017low}
S.~Lankes, S.~Pickartz, J.~Breitbart, A low noise unikernel for extrem-scale
  systems, in: International Conference on Architecture of Computing Systems,
  Springer, 2017, pp. 73--84.

\bibitem{hourglass}
G.~Wassen, Hourglass benchmark github repository,
  \url{https://github.com/georgwassen/hourglass}.

\bibitem{regehr2002inferring}
J.~Regehr, Inferring scheduling behavior with hourglass., in: USENIX Annual
  Technical Conference, FREENIX Track, 2002, pp. 143--156.

\bibitem{klein2014comprehensive}
G.~Klein, J.~Andronick, K.~Elphinstone, T.~Murray, T.~Sewell, R.~Kolanski,
  G.~Heiser, Comprehensive formal verification of an os microkernel, ACM TOCS
  32~(1) (2014) 1--70.

\bibitem{kohler2000click}
E.~Kohler, R.~Morris, B.~Chen, J.~Jannotti, M.~F. Kaashoek, The click modular
  router, ACM TOCS 18~(3) (2000) 263--297.

\bibitem{hermes_project}
{HERMES2020}, {qualification of High pErformance pRogrammable Microprocessor
  and dEvelopment of Software ecosystem},
  \url{https://cordis.europa.eu/project/id/101004203}.

\bibitem{dom0less_1}
{The Linux Foundation}, {True Static Partitioning with Xen Dom0-less},
  \url{https://xenproject.org/2019/12/16/true-static-partitioning-with-xen-dom0-less/}.

\bibitem{dom0less_2}
{The Linux Foundation}, {Dom0less},
  \url{https://xenbits.xen.org/docs/4.15-testing/features/dom0less.html}.

\bibitem{colp2011breaking}
P.~Colp, M.~Nanavati, J.~Zhu, W.~Aiello, G.~Coker, T.~Deegan, P.~Loscocco,
  A.~Warfield, Breaking up is hard to do: security and functionality in a
  commodity hypervisor, in: Proceedings of the Twenty-Third ACM Symposium on
  Operating Systems Principles, 2011, pp. 189--202.

\bibitem{dom0_disaggregation}
{The Linux Foundation}, {Dom0 Disaggregation},
  \url{https://wiki.xenproject.org/wiki/Dom0_Disaggregation}.

\bibitem{mvondo2020fine}
D.~Mvondo, A.~Tchana, R.~Lachaize, D.~Hagimont, N.~De~Palma, Fine-grained fault
  tolerance for resilient pvm-based virtual machine monitors, in: Proc. DSN,
  IEEE, 2020, pp. 197--208.

\bibitem{elisa_project}
{The Linux Foundation}, {Homepage of Enabling Linux In Safety Applications
  (ELISA) project}, \url{https://elisa.tech/}.

\bibitem{lf_edge}
{The Linux Foundation}, {Homepage of LF Edge Foundation},
  \url{https://elisa.tech/}.

\bibitem{li2019acrn}
H.~Li, X.~Xu, J.~Ren, Y.~Dong, Acrn: a big little hypervisor for iot
  development, in: Proc. VEE, ACM, 2019, pp. 31--44.

\bibitem{morabito2018consolidate}
R.~Morabito, V.~Cozzolino, A.~Y. Ding, N.~Beijar, J.~Ott, Consolidate iot edge
  computing with lightweight virtualization, IEEE Network 32~(1) (2018)
  102--111.

\bibitem{goldschmidt2018container}
T.~Goldschmidt, S.~Hauck-Stattelmann, S.~Malakuti, S.~Gr{\"u}ner,
  Container-based architecture for flexible industrial control applications,
  Elsevier JSA 84 (2018) 28--36.

\bibitem{morabito2017evaluating}
R.~Morabito, I.~Farris, A.~Iera, T.~Taleb, Evaluating performance of
  containerized iot services for clustered devices at the network edge, IEEE
  Internet of Things Journal 4~(4) (2017) 1019--1030.

\bibitem{kurek2019unikernel}
T.~Kurek, Unikernel network functions: A journey beyond the containers, IEEE
  Communications Magazine 57~(12) (2019) 15--19.

\bibitem{iso24029}
{ISO}, {Assessment of the robustness of neural networks — Part 1: Overview},
  ISO/IEC DTR 24029-1: Artificial Intelligence (AI).

\bibitem{iso4213}
{ISO}, {Assessment of machine learning classification performance}, ISO/IEC WD
  TS 4213 Information technology — Artificial Intelligence.

\end{thebibliography}

\end{document}